\documentclass[12pt,preprint]{aastex}

\slugcomment{}

\shorttitle{Photometric selection of high-redshift type Ia supernovae}
\shortauthors{Sullivan et al.}

\newcommand\omatter{\ifmmode \Omega_{M}\else $\Omega_{M}$\fi}
\newcommand\ok{\ifmmode \Omega_{\mathrm{k}}\else $\Omega_{\mathrm{k}}$\fi}
\newcommand\olambda{\ifmmode \Omega_{\Lambda}\else $\Omega_{\Lambda}$\fi}
\newcommand\dmB{\ifmmode \Delta m_{15}(B) \else $\Delta m_{15}(B)$\fi}
\newcommand\zspec{\ifmmode z_{\mathrm{spec}}\else $z_{\mathrm{spec}}$\fi}
\newcommand\zphot{\ifmmode z_{\mathrm{phot}}\else $z_{\mathrm{phot}}$\fi}
\newcommand\deltaz{\ifmmode \Delta z \else $\Delta z$\fi}
\newcommand\chidof{\ifmmode \chi^2/\mathrm{DOF}\else $\chi^2/\mathrm{DOF}$\fi}
\newcommand\ebmvmw{\ifmmode E_{B-V}^{\small \mathrm{mw}}\else $E_{B-V}^{\small \mathrm{mw}}$\fi}
\newcommand\ebmvhost{\ifmmode E_{B-V}^{\small \mathrm{host}}\else $E_{B-V}^{\small \mathrm{host}}$\fi}

\begin{document}

\title{Photometric selection of high-redshift Type Ia supernova candidates}

\author{M.~Sullivan\altaffilmark{1}, D.~A.~Howell\altaffilmark{1}, K.~Perrett\altaffilmark{1}, P.~E.~Nugent\altaffilmark{2}, P.~Astier\altaffilmark{3}, E.~Aubourg\altaffilmark{4,5}, D.~Balam\altaffilmark{6}, S.~Basa\altaffilmark{7}, R.~G.~Carlberg\altaffilmark{1}, A.~Conley\altaffilmark{1}, S.~Fabbro\altaffilmark{8}, D.~Fouchez\altaffilmark{9}, J.~Guy\altaffilmark{3}, I.~Hook\altaffilmark{10}, H.~Lafoux\altaffilmark{5}, J.~D.~Neill\altaffilmark{6}, R.~Pain\altaffilmark{3}, N.~Palanque-Delabrouille\altaffilmark{5}, C.~J.~Pritchet\altaffilmark{6}, N.~Regnault\altaffilmark{3}, J.~Rich\altaffilmark{5}, R.~Taillet\altaffilmark{11,3}, G.~Aldering\altaffilmark{12}, S.~Baumont\altaffilmark{3}, J.~Bronder\altaffilmark{10}, M.~Filiol\altaffilmark{7}, R.~A.~Knop\altaffilmark{13}, S.~Perlmutter\altaffilmark{12}, C.~Tao\altaffilmark{9}}

\altaffiltext{1}{Department of Astronomy and Astrophysics, University of Toronto, 60 St. George Street, Toronto, ON M5S 3H8, Canada}
\altaffiltext{2}{Computational Research Division, Lawrence Berkeley National Laboratory, MS 50F-1650, 1 Cyclotron Rd, Berkeley, CA 94720, USA}
\altaffiltext{3}{LPNHE, CNRS-IN2P3 and University of Paris VI \& VII, 75005 Paris, France}
\altaffiltext{4}{APC, 11 Pl. M. Berthelot, 75231 Paris Cedex 5, France}
\altaffiltext{5}{DSM/DAPNIA, CEA/Saclay, 91191 Gif-sur-Yvette Cedex, France}
\altaffiltext{6}{Department of Physics and Astronomy, University of Victoria, PO Box 3055, Victoria, BC V8W 3P6, Canada}
\altaffiltext{7}{LAM CNRS, BP8, Traverse du Siphon, 13376 Marseille Cedex 12, France}
\altaffiltext{8}{CENTRA - Centro Multidisciplinar de Astrof\'{\i}sica, IST, Avenida Rovisco Pais, 1049 Lisbon, Portugal}
\altaffiltext{9}{CPPM, CNRS-IN2P3 and University Aix Marseille II, Case 907, 13288 Marseille Cedex 9, France}
\altaffiltext{10}{University of Oxford Astrophysics, Denys Wilkinson Building, Keble Road, Oxford OX1 3RH, UK}
\altaffiltext{11}{Université de Savoie, 73000 Chamb\'ery, France}
\altaffiltext{12}{Lawrence Berkeley National Laboratory, 1 Cyclotron Rd., Berkeley, CA 94720, USA}
\altaffiltext{13}{Department of Physics and Astronomy, Vanderbilt University, VU Station B 351807, Nashville, TN 37235-1807 USA}

\begin{abstract}
  
We present a method for selecting high-redshift type Ia supernovae
(SNe~Ia) located via rolling SN searches. The technique, using both
color and magnitude information of events from only 2-3 epochs of
multi-band real-time photometry, is able to discriminate between
SNe~Ia and core collapse SNe. Furthermore, for the SNe~Ia, the
method accurately predicts the redshift, phase and light-curve
parameterization of these events based only on pre-maximum-light
data. We demonstrate the effectiveness of the technique on a
simulated survey of SNe~Ia and core-collapse SNe, where the
selection method effectively rejects most core-collapse SNe while
retaining SNe~Ia.  We also apply the selection code to real-time
data acquired as part of the Canada-France-Hawaii Telescope
Supernova Legacy Survey (SNLS). During the period May 2004 to
January 2005 in the SNLS, 440 SN candidates were discovered of which
70 were confirmed spectroscopically as SNe~Ia and 15 as
core-collapse events. For this test dataset, the selection technique
correctly identifies 100\% of the identified SNe~II as non-SNe~Ia
with only a 1-2\% false rejection rate. The predicted
parameterization of the SNe~Ia has a precision of
$\left|\deltaz\right|/(1+\zspec)<0.09$ in redshift, and $\pm$2-3
rest-frame days in phase, providing invaluable information for
planning spectroscopic follow-up observations. We also investigate
any bias introduced by this selection method on the ability of
surveys such as SNLS to measure cosmological parameters (e.g., $w$
and \omatter), and find any effect to be negligible.

\end{abstract}

\keywords{surveys -- supernovae: general -- distance scale}

\section{Introduction}
\label{sec:introduction}

Recent surveys for high-redshift type Ia supernovae (SNe~Ia) have
established their utility as calibrated standard candles suitable for
use as cosmological distance indicators.  The ``first generation'' of
SN~Ia cosmology programs
\citep{1997ApJ...483..565P,1998Natur.391...51P,1998ApJ...493L..53G,1998AJ....116.1009R,1998ApJ...507...46S,1999ApJ...517..565P}
developed a systematic approach to high-redshift SN detection and
analysis that led to Hubble diagrams ruling out a flat,
matter-dominated Universe, revealing the presence of an
unaccounted-for ``dark energy'' driving cosmic acceleration.

One of the most pressing questions in cosmology now is: ``What is the
nature of the dark energy?''. There is a fundamental difference
between a Cosmological Constant and other proposed forms of dark
energy, and the distinction can be addressed by measuring the dark
energy's average equation-of-state, $w\equiv p/\rho$, where $w=-1$
corresponds to a Cosmological Constant. Current measurements of this
parameter
\citep[e.g.][]{2003ApJ...598..102K,2003ApJ...594....1T,2004ApJ...607..665R}
are consistent with a very wide range of dark energy theories
\citep[see][for a comprehensive review]{2003RvMP...75..559P}. The
importance of improving measurements to the point where $w=-1$ could
be confirmed or excluded has led to a ``second-generation'' of
supernova cosmology studies: large multi-year, multi-observatory
programs benefiting from major commitments of dedicated time, e.g. the
Supernova Legacy Survey at the Canada-France-Hawaii Telescope
\citep{2005astro.ph.10447A}, and the ESSENCE project at the Cerro
Tololo Inter-American Observatory \citep{2005astro.ph..8681K}.  These
``rolling searches'' find and follow all varieties of SNe over many
consecutive months of repeated wide-field imaging, with redshifts and
SN type classification from coordinated spectroscopy. The success of
these large-scale rolling searches directly depends on the degree to
which SNe~Ia can be selected at early phases in their light-curve from
many other types of varying sources for further spectroscopic
follow-up, and hence confirmation of their nature. With large numbers
of candidates being found during each lunation, and a finite amount of
8m-class telescope spectroscopic follow-up available, these
spectroscopy-limited surveys must have reliable methods for extracting
likely SN~Ia events from the hundreds of other varying sources
discovered.

Previous studies have investigated methods by which SN can be
identified from light-curve or host-galaxy photometric properties
alone. \citet{2002PASP..114..833P} used color-color diagrams and SN
spectral templates at various redshifts to delineate regions of
color-color space in which SNe of different types are usually located;
a particular adaptation of this technique is to identify SNe~Ib/c
\citep{2004PASP..116..597G}.  \citet{2002PASP..114..284D} demonstrated
that photometric properties of the host galaxies and SNe can be used
to select high-redshift candidates in well-defined redshift intervals
for spectroscopic follow-up. \citet{2004ApJ...600L.163R} used color
cuts to differentiate SNe~Ia at $z>1$ from lower-redshift SNe~II,
while \citet{2004ApJ...602..571B} used photometric techniques to
attempt to identify high-redshift SN candidates (with complete
light-curves) as SNe~Ia based on a spectroscopic redshift.
\citet{2004ApJ...613L..21B} introduced a redshift marginalization
technique which allowed SNe~Ia to be used without knowledge of a
spectroscopic redshift with an impressively small increase in scatter
about the $z<0.13$ Hubble diagram; however this technique used full
light-curves and an already obtained positive identification of the SN
type.

Many (though not all) of these techniques are designed to identify SNe
either after a host galaxy spectroscopic redshift has been determined,
or after a full light-curve has been obtained. The problem facing
rolling-searches is different, namely differentiating SNe~Ia from all
other variable sources (e.g., SNe~II, AGN, variable stars) based on
only 2-3 epochs of real-time (i.e.  perhaps imperfect) photometry, and
predicting their redshifts, phases, and temporal evolution in
magnitude. Redshift estimates allow spectrograph setups to be
optimized (in terms of grism or grating central wavelength choice) for
the candidate under study, to ensure that the characteristic spectral
features of SNe~Ia are always within the observed wavelength range to
allow unambiguous identification. Phase predictions allow
spectroscopic follow-up to be optimized by always observing SN
candidates within a few days of maximum light, when the SNe have the
greatest contrast over their host galaxies, and when spectral features
in different types of SNe are most easily distinguished.  Predicting
the magnitude of a SN over the course of its light-curve is invaluable
in ensuring that spectroscopic observations are of a sufficient length
to obtain a signal-to-noise (S/N) that allows an identification to be
obtained.

In this paper, we present a new technique which is designed to select
and predict the redshift, phase and light-curve parameterization of
high-redshift SNe~Ia. We show how it has been applied to real-time
data from the Supernova Legacy Survey (SNLS) and used to guide and
inform follow-up decisions.  A plan of the paper follows.
$\S$\ref{sec:photometric-selection} gives a brief overview of SN~Ia
properties and describes the technique used for our photometric
selection; $\S$\ref{sec:fakedata} shows the results of our tests of
the technique on synthetic datasets of SNe~Ia and core collapse SNe.
The technique is applied to candidates from the SNLS in
$\S$~\ref{sec:photometric-selection-application}, the results of which
are presented, compared with simulations, and discussed in
$\S$\ref{sec:results}. We summarise in $\S$\ref{sec:summary}.
Throughout this paper all magnitudes are given in the AB photometric
system \citep{1983ApJ...266..713O}, and we assume a cosmology of
$\omatter=0.30$, $\olambda=0.70$, and $w=-1$ where relevant.

\section{Photometric selection technique}
\label{sec:photometric-selection}

Given flux information about a candidate SN, we require a method that
provides a light-curve fit and parameterization of the SN (assuming
the candidate is a SN~Ia), an estimate of the SN redshift, and some
quality factor providing a measure of the likelihood that the
candidate under study is indeed a SN~Ia.

\subsection{Parameterizing Type Ia supernovae light-curves}
\label{sec:type-ia-supernovae}

Raw observed peak magnitudes of spectroscopically normal SNe~Ia have
an r.m.s. dispersion of $\sim0.30\,\mathrm{mag}$
\citep[e.g.][]{1996AJ....112.2391H}, but this dispersion can be
significantly reduced by using an empirical relationship between the
light-curve shape and the raw observed peak magnitude.
\citet{1993ApJ...413L.105P} showed that the absolute magnitudes of
SNe~Ia are tightly correlated with the initial decline rate of the
light-curve in the rest-frame $B$-band -- intrinsically brighter SNe
have wider, more slowly declining light-curves. Using a linear
relationship to correct the observed SN peak magnitudes, parameterized
via \dmB, the decline in $B$-band magnitudes 15 days after maximum
light, the typical scatter in the rest-frame $B$-band of local SNe can
be reduced to $\sim0.17\,\mathrm{mag}$
\citep{1995AJ....109....1H,1996AJ....112.2391H,1999AJ....118.1766P}.
An alternative technique, the Multi-color Light Curve Shape
\citep[MLCS;][]{1995ApJ...438L..17R,1996ApJ...473...88R,2004ApJ...607..665R,2002PhDT........10J},
results in a similar smaller dispersion in the final peak magnitudes,
particularly if multi-color light-curves are utilized in the fitting
procedure.

The approach used in this paper is to parameterize the SN light-curve
time-scale via a simple stretch factor, $s$, which linearly stretches
or contracts the time axis of a template SN light curve around the
time of maximum light to best-fit the observed light curve of the SN
being fit
\citep[][]{1997ApJ...483..565P,1999ApJ...517..565P,2001ApJ...558..359G}.
A SN's `corrected' peak magnitude ($m^{\mathrm{corr}}_{B}$) is then
related to its raw peak magnitude ($m^{\mathrm{raw}}_{B}$) via

\begin{equation}
\label{eq:1}
  m^{\mathrm{corr}}_{B}=m^{\mathrm{raw}}_{B}+\alpha(s-1)
\end{equation}

\noindent
where $s$ is the stretch of the light-curve time-scale and $\alpha$
relates $s$ to the size of the magnitude correction. Sophisticated
versions of this technique incorporating color corrections
\citep[e.g.][]{2005guy} lead to r.m.s.  dispersions of
$\simeq0.15$\,mag in local SNe.

\subsection{SN model}
\label{sec:SN-model}

The apparent observed flux $f$ of a SN~Ia at redshift $z$ through a
filter with total system response $T(\lambda)$ is given by

\begin{equation}
\label{eq:2}
f_T=\frac{1}{1+z}\int_0^{\infty} T(\lambda) S\left(\frac{\lambda}{1+z},t,s,\ebmvmw,\ebmvhost\right) \mathrm{d}\lambda
\end{equation}

\noindent
where $S(\lambda,t,s,\ebmvmw,\ebmvhost)$,
hereafter $S$, is the rest-frame flux distribution of the SN (in
erg\,s$^{-1}$\,cm$^{-2}$\,\AA$^{-1}$), which depends on the phase of the SN
relative to maximum light ($t$, measured in rest-frame days), the SN
stretch ($s$), the color excess due to the Milky Way along the
line-of-sight to the SN (\ebmvmw), and the color excess
of the SN host galaxy (\ebmvhost).  $S$ is normalized so
that its intrinsic, unextincted, $B$-band magnitude at maximum light
($t=0$) $m_{B}^{\mathrm{raw}}$ is given by

\begin{equation}
\label{eq:3}
  m_{B}^{\mathrm{raw}}={\cal M}_{B}+5\log {\cal D}_{\cal{L}}(z;\omatter,\olambda)-\alpha(s-1)
\end{equation}

\noindent
where ${\cal D}_{\cal L} \equiv H_0 d_L$ is the ``Hubble
constant-free'' luminosity distance, and ${\cal M}_B\equiv M_B-5\log
H_0+25$ is the ``Hubble constant-free'' $B$-band absolute peak
magnitude of a SN~Ia with $s=1$
\citep[e.g.][]{1995ApJ...450...14G,1999ApJ...517..565P}. In this
model, we set $\alpha=1.47$ and ${\cal M}_{B}=-3.48$ from the primary
fit of \citet{2003ApJ...598..102K}. (This formalism can of course be
generalized to any SN type by removing the dependence on the stretch,
$s$.)

Model spectra and light-curves for the SNe are also required.  The
basic, normal SN~Ia spectral templates we use are updated versions of
those given in \citet{2002PASP..114..803N}, which provide spectral
coverage from 1000\AA\ to 25000\AA\ in wavelength, and from $t=-19$ to
$t=+70$ rest-frame days in the SN light-curve. The light-curve
templates for calculating rest-frame $B$-band fluxes relative to
maximum light (before any stretch-color correction) are the rest-frame
light-curve templates of \citet{2001ApJ...558..359G} and
\citet{2003ApJ...598..102K}, linearly interpolating between days as
required. Light-curve information is provided for $t=-19$ to $t=+330$
rest-frame days; for epochs before $t=-19$ days, the SN flux is
assumed to be zero.

We also incorporate the observed relationship between stretch and
color from \citet{2003ApJ...598..102K}, which describes the variation
in SN rest-frame $U-B$, $B-V$, $V-R$ and $R-I$ colors as a function of
stretch. For a given epoch, the spectral template is adjusted to have
the colors for a given stretch using a spline which smoothly scales
the spectrum as a function of wavelength.  Extinction is applied to
the altered template using the reddening law of
\citet{1989ApJ...345..245C} updated in the near-UV by
\citet{1994ApJ...422..158O}, with $R_V=3.1$, first on the rest-frame
spectrum with \ebmvhost, and then on the redshifted spectrum with
\ebmvmw. We use the filter responses of \citet{1990PASP..102.1181B}
and the $\alpha$-Lyrae spectra energy distribution of
\citet{2004AJ....127.3508B} to perform a conversion to the AB system
when required.

\subsection{Fitting technique}
\label{sec:Fitting-technique}

For each SN candidate, we need to perform a fit to the $n$ observed
flux values $O$ with uncertainty $O_{\mathrm{err}}$ which depend on
the filter $T$ used for the observation, and the time of the
observation $t_i$. We calculate how well a given model matches the
observed data using the $\chi^2$ statistic, i.e.

\begin{equation}
\label{eq:4}
\chi^2=\sum_{i=1}^n\left(\frac{(O_i-f_T\left(T_i;z,t_i-t_{max},s,\ebmvmw,\ebmvhost\right))}{O_i^{\mathrm{err}}}\right)^2
\end{equation}

\noindent
where $f_T$ is defined in equation~(\ref{eq:2}). Our task is to derive
the model parameters $z$, $s$, $t_{max}$, \ebmvhost\ which best fit
the data, i.e. which minimize $\chi^2$. While a grid search in
$\chi^2$-space is possible, in practice, with the number of free
parameters in the model and typically a large number of potential SNe
to fit, such a procedure becomes too time consuming for ``real-time''
operation.

Instead, the best-fitting model templates are located using
MPFIT\footnote{http://cow.physics.wisc.edu/\~{}craigm/idl/idl.html}, a
robust non-linear least-squares fitting program written in the IDL
programming language translated from the MINPACK-1 FORTRAN package
\citep{minpack}.  \ebmvmw\ is taken from the dust maps of
\citet{1998ApJ...500..525S} and held fixed in the model. The output
from this procedure are the best-fitting parameters of redshift,
stretch and epoch, and optionally \ebmvhost. An additional parameter,
termed a ``dispersion magnitude'' ($\mathrm{d}m$) which is added to or
subtracted from the effective $B$-band magnitude of a model SN, is
also fit. Physically, this accounts for the dispersion in
light-curve-shape corrected peak magnitudes seen in local SN samples,
of $\sim0.17$\,mag; in our fits, this value is allowed to vary from
-0.2 to 0.2\,mag, but is not used further in the analysis. The other
fit parameters are allowed to vary over the following ranges:
$0.6<s<1.4$, $0.05<z<1.4$, and $-0.10<\ebmvhost<0.30$, representing
the typical parameter range of SNe~Ia found in rolling searches.

Once a first fit has been performed, the observed data are trimmed to
only include data lying in the rest-frame to +30 days, the epoch to
which stretch is most effective \citep{2001ApJ...558..359G}. Once a
first redshift estimate is made, data in an observed filter which
corresponds to rest-frame $R$ or redder (where the current stretch
technique is known to have limitations) are excluded from future
iterations. For the Sloan Digital Sky Survey (SDSS) filter system,
this implies that if the predicted redshift is less than 0.50, $z'$
data are excluded from the second iteration fit.

\section{Testing on simulated survey data}
\label{sec:fakedata}

We now test the performance of the SN selection technique described
above on simulated datasets, as a consistency check that our technique
can successfully recover the input parameters of fake SNe~Ia and
determine their properties without any bias, and that a basic
discrimination between SNe~Ia and core-collapse SNe can be made. We
generate both SNe~Ia and core-collapse SNe in this simulation
according to the design of the SNLS.

\subsection{SNLS survey design}
\label{sec:snls-supern-legacy}

The Canada-France-Hawaii Telescope Legacy Survey (CFHT-LS) began in
June 2003 using the queue-scheduled wide-field imager ``Megacam'' on
the CFHT on Mauna Kea.  Megacam consists of 36 2048$\times$4612 pixel
CCDs covering a 1\degr $\times$1\degr\ field-of-view. CFHT-LS
comprises three separate surveys (Deep, Wide, and Very Wide), with the
SNLS exploiting data from the Deep component. This component of the
survey conducts repeat observations of 4 low-extinction fields
distributed in right ascension (Table~\ref{tab:fieldpositions}).

In a typical month, each available field is imaged on five epochs in a
combination of $g'r'i'z'$ filters which closely match those used for
the photometric calibration of the SDSS.  $r'i'$ observations are
always taken, and $g'z'$ observations are arranged according to the
lunar phase. The typical break-down in exposure times for each field
is given in Table~\ref{tab:exposure_times}; in general, each of the
epochs are spaced 4-5 days apart ($\sim$3 days in a typical SN
rest-frame).  ($u'$ data of the fields are also taken, but are not
time sequenced as they are not used for the SN light-curves.)  Each
field is typically observed for 4-5 continuous months, plus a final
month (with fewer epochs) to complete light-curves, resulting in
around 20 ``field-months'' in a calendar year. The queue-scheduled
nature of the observations provides a powerful protection against
inclement weather, with the result that very few epochs are totally
lost -- most months have received the full allocation of 5 observation
epochs per field.  Consequently high-quality, multi-filter, continuous
light-curves for each SN candidate lasting many months are routinely
obtained.

\subsection{Simulating the SNLS}
\label{sec:simulating-snls}

We generate synthetic datasets as follows. For SNe~Ia, events are
drawn randomly from a population with truncated Gaussian distributions
in redshift ($\overline{z}=0.6$, $\sigma=0.4$,
$z_{\mathrm{min}}=0.15$, $z_{\mathrm{max}}=1.05$), stretch
($\overline{s}=1$, $\sigma=0.25$, $s_{\mathrm{min}}=0.6$,
$s_{\mathrm{max}}=1.4$), dispersion in peak magnitude
($\overline{\mathrm{d}m}=0$, $\sigma=0.17$), and host galaxy
extinction ($\overline{\ebmvhost}=0$, $\sigma=0.2$, $\ebmvhost>-0.1$).

In contrast to SNe~Ia, core-collapse SNe are quite heterogeneous
events, with a large variation in their light-curves, spectra, colors
and peak magnitudes; currently there is no well-established unifying
scheme (c.f. stretch or \dmB\ for SN~Ia) for the spectral and
light-curve observations \citep[though hints of a relationship between
light-curve shape and luminosity have been observed for
SNe~Ic;][]{2005ApJ...626L...5S}. Nonetheless, we attempt to simulate
them here using what little data is currently available in the
literature, with the caveat that our simulations may not represent the
full range of core-collapse SNe that could be observed. 

We generate four sub-types of core-collapse SNe, all based on spectral
templates constructed by one of us (PEN). For SN~Ib/c, we use the
spectral templates\footnote{The templates can be obtained at
  http://supernova.lbl.gov/\~{}nugent/nugent\_templates.html}
described in \citet{2005ApJ...624..880L}, and our spectral templates
for SN~IIL and SN~IIP are similar to those used in
\citet{1999ApJ...521...30G}. We also created a SN~IIn template based
on SN~1999el \citep{2002ApJ...573..144D}. The peak absolute magnitudes
and dispersions for the core-collapse SNe are taken from
\citet{2002AJ....123..745R}; Gaussian distributions are again assumed.
The other parameters defining the core collapse population are
truncated Gaussians: redshift ($\overline{z}=0.5$, $\sigma=0.3$,
$z_{\mathrm{min}}=0.05$, $z_{\mathrm{max}}=0.6$), and host galaxy
extinction ($\overline{\ebmvhost}=0$, $\sigma=0.3$, $\ebmvhost>-0.2$).
These values differ from those used for the SNe~Ia, and reflect their
intrinsic faintness when compared to SNe~Ia, and likely increased dust
extinction on these events (due to their association with recent
star-formation).

The simulations use an observational cadence similar to that of the
SNLS i.e. 5 observation epochs spread over 16 day-long observing
periods centered around new moon (Table~\ref{tab:exposure_times}). Any
given epoch has a 15\% chance of being totally lost due to external
factors, such as long periods of adverse weather. For each of these
epochs, the rest-frame template SN spectrum is adjusted for the
stretch-color relationship (for SNe~Ia only), and additional
dispersions in SN color are added, accounting for the natural
variation in SN properties. For SNe~Ia, the sigma of the dispersions
used are 0.08\,mag for $U-B$, and 0.03\,mag for $B-V$, $V-R$ and
$R-I$. These dispersions are those intrinsic to SNe~Ia after
stretch-color relations have been applied. The increased dispersion in
$U-B$ over the other colors reflects the increasing sensitivity of the
bluer passbands to chemical abundances
\citep[e.g.][]{1998ApJ...495..617H,2000ApJ...528..590H,2000ApJ...530..966L},
and represent the broad range of observed values found in the
literature
\citep{2003A&A...404..901N,2003ApJ...598..102K,2005astro.ph..9234J}.
For core-collapse SNe, the dispersion we use are: 0.20\,mag for $U-B$,
and 0.15\,mag for $B-V$, $V-R$ and $R-I$. The increased values are
deliberately conservative, and represent our relatively poor knowledge
of core-collapse SNe.

\begin{figure}
\plotone{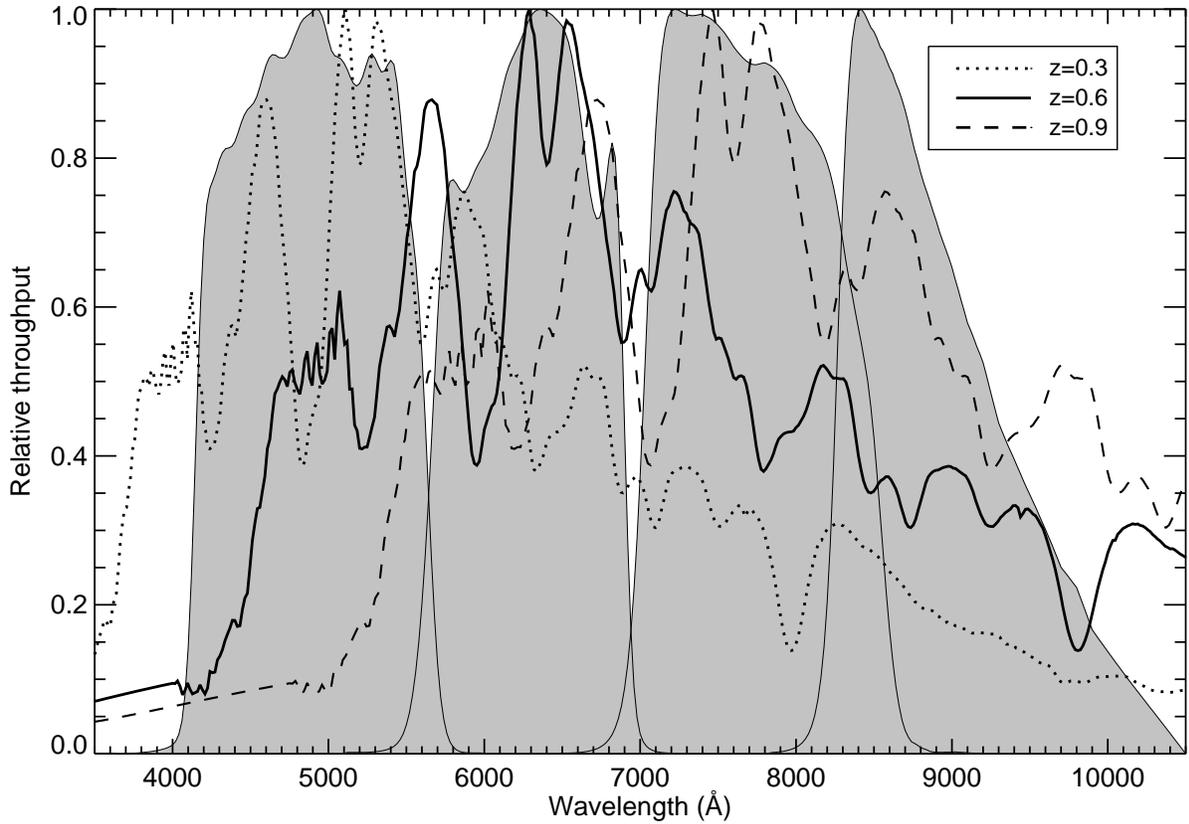}
\caption{
  The mean effective responses of the $g'$, $r'$, $i'$, and $z'$
  filters used in CFHT-LS, shown shaded left to right in the figure. The
  filter responses are those measured by CFHT, additionally multiplied
  by the reflectivity of the primary mirror, the throughput of the
  wide-field corrector optics, and the typical Megacam CCD quantum
  efficiency.  Plotted underneath is a maximum-light $s=1$ SN~Ia
  spectrum at redshifts of 0.3 (dotted), 0.6 (solid) and 0.9 (dashed),
  spanning the range of SNe~Ia followed by the SNLS.  The SN template
  is a revised version from \citet{2002PASP..114..803N}.
\label{fig:filters}
}

\end{figure}

The SN spectrum is then scaled to the expected effective $B$-band
magnitude for that phase (assuming an $\omatter=0.30$, $\olambda=0.70$
cosmology), the stretch-luminosity relationship and the dispersion in
peak magnitude from $\mathrm{d}m$ are applied, host extinction is
applied, the spectrum is redshifted, and finally Milky Way extinction
is applied. The spectrum is then integrated through the SNLS filters
to generate an apparent magnitude in each filter, and the Megacam
zeropoints used to calculate the counts in that filter given the
exposure times listed in Table~\ref{tab:exposure_times}. The Megacam
filter responses are those measured and provided by CFHT, and are
additionally multiplied by the reflectivity of the CFHT primary
mirror, the throughput of the wide-field corrector plus optics, and
the average Megacam CCD quantum efficiency. These effective filter
responses can be seen in Figure~\ref{fig:filters}, together with a
typical SN~Ia spectrum at a range of redshifts typical of SNLS
candidates.

Poisson noise, a filter-dependent sky noise (measured from real data
in dark and gray sky conditions), and a filter-dependent systematic
``flat-field error'' noise is added to the predicted counts, and the
counts converted back into fluxes using the original zeropoint, but
with an additional zeropoint uncertainty ($\sigma=0.05$\,mag) applied,
typical of that present in \textit{real-time} SNLS data \citep[the
zeropoint uncertainty for the final photometry is 0.02\,mag or less,
see][]{2005astro.ph.10447A}. Additional systematic noise is added to the $z'$
(and a smaller amount in $i'$) data to account for the imperfect
fringe correction in real-time data.  The result is a set of
observations which should closely match real SNLS data.  These
synthetic SN observations are then passed to the SN selection code,
and the code is run to locate the best-fitting SN parameters. This
experiment was repeated for 2000 synthetic SNe. The final population
of SNe has the following percentage composition: SNe~Ia 50\%, SN~Ib/c
25\%, SN~IIL 8\%, SN~IIP 8\%, and SN~IIn 8\%.

\subsection{Simulation Results for SNe~Ia}
\label{sec:simulation-results}

\begin{figure} 
\plotone{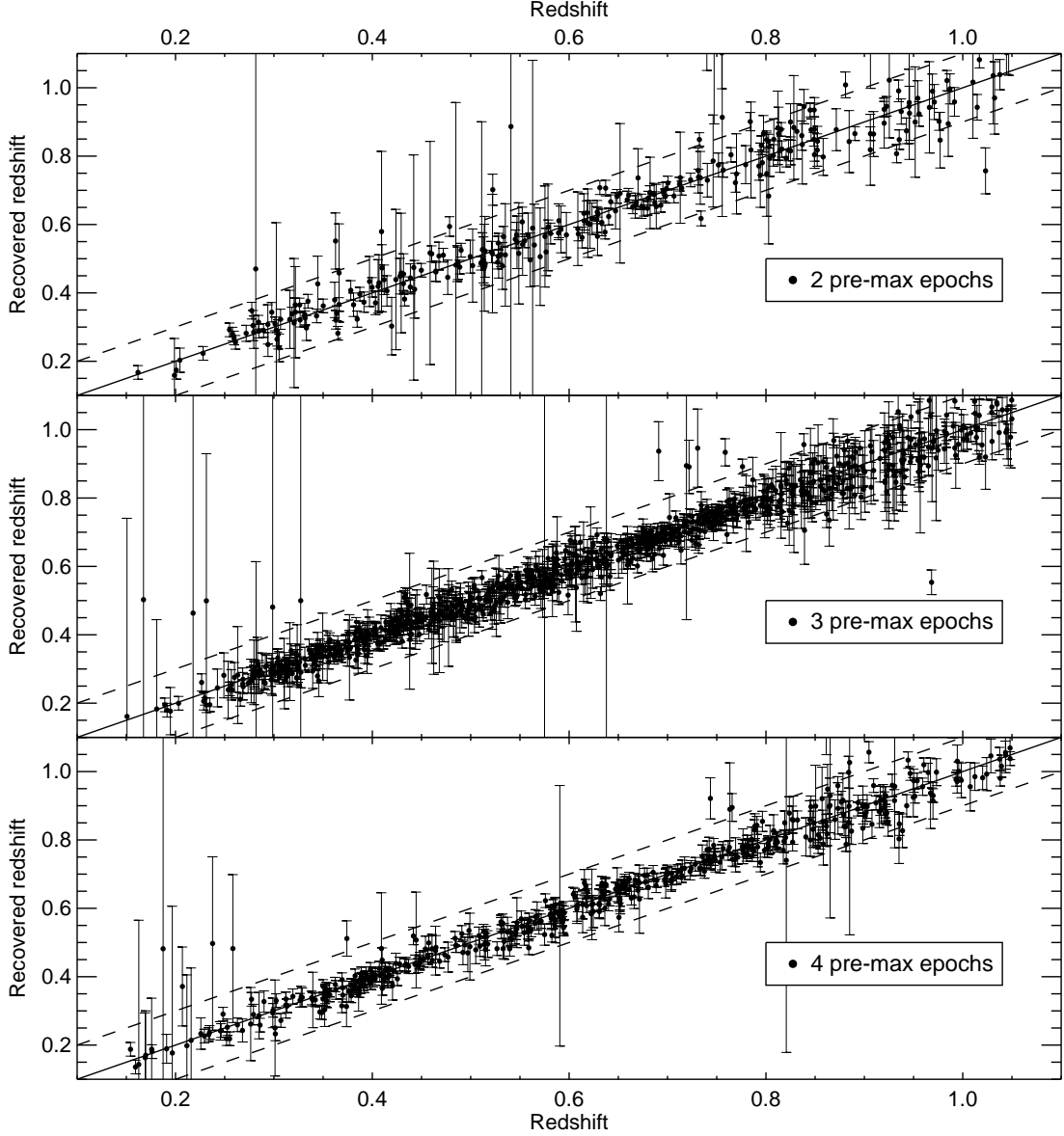}
\caption{
  Results of the SN selection code run on a synthetic dataset of 2000
  SNe~Ia generated with observational parameters similar to that found
  in the SNLS (see $\S$~\ref{sec:simulating-snls} for simulation
  details). This figure shows the comparison of actual and recovered
  redshift, and how this varies depending on the number of pre-maximum
  light-curve epochs are used in the fit, excluding pre-SN explosion
  epochs, from 2 epochs (top panel) to 4 epochs (lower panel). Only
  pre-maximum light data (including epochs of non detections) are used
  when recovering the redshift. On each epoch, 3 filters are observed,
  either $g'r'i'$ or $r'i'z'$.
  \label{fig:synthetic-alldata-z}}
\end{figure}

\begin{figure} 
\plotone{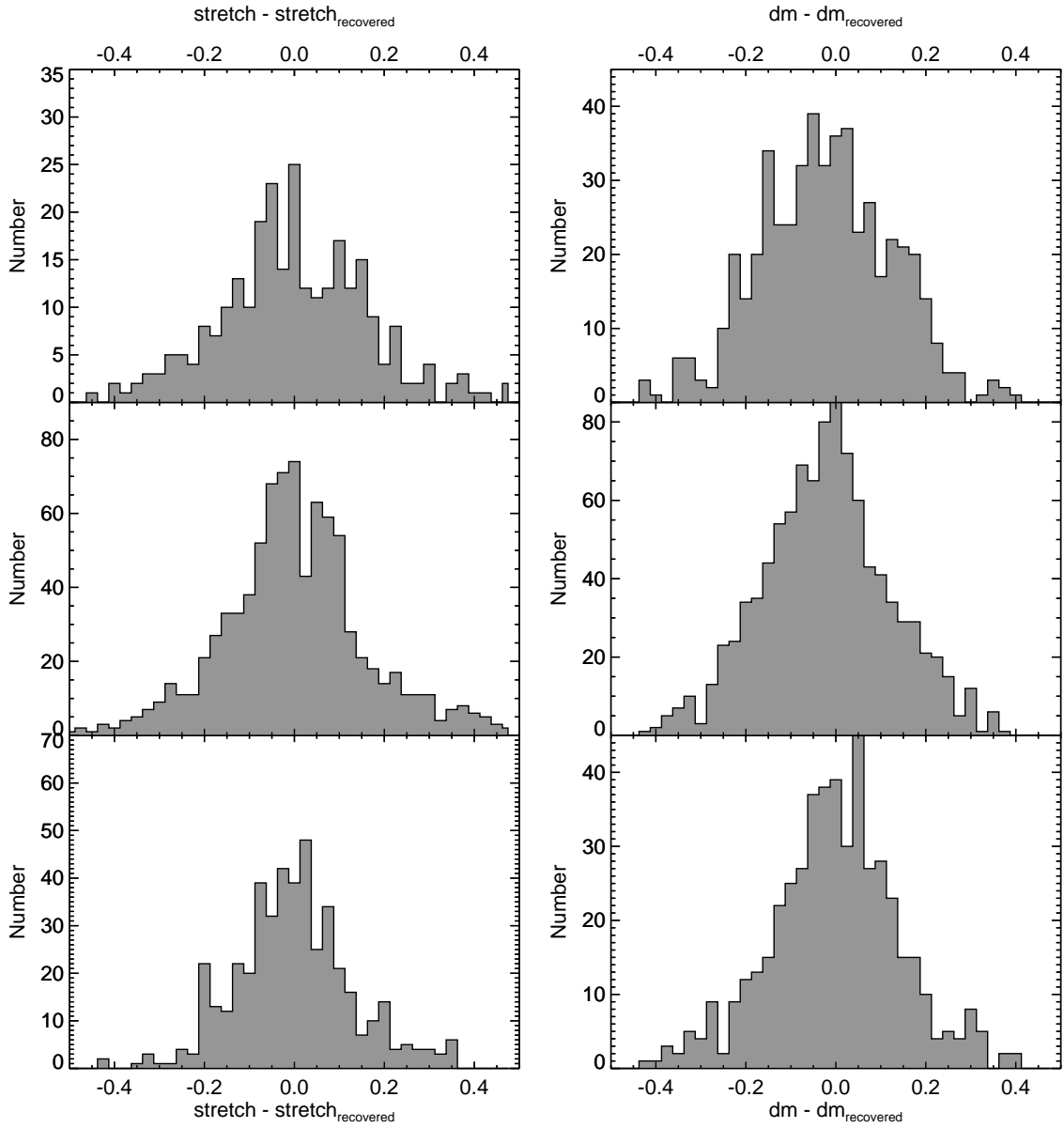}
\caption{
  As Figure~\ref{fig:synthetic-alldata-z}, but showing the distribution
  of actual minus recovered stretch (left panels), and for
  $\mathrm{d}m$ (right panels).
  \label{fig:synthetic-alldata-sdm}}
\end{figure}

We now examine the results of the fits to the simulated SN~Ia
photometry (the core collapse SNe are discussed in
Section~\ref{sec:results}).  In Figures~\ref{fig:synthetic-alldata-z}
and \ref{fig:synthetic-alldata-sdm} we compare the results of the
input and recovered SN~Ia parameters. This figure shows the comparison
of actual versus recovered redshift, and the difference in the
recovered stretch and $\mathrm{d}m$ parameters.  We also show how the
accuracy of the recovered parameters depends on the number of data
epochs in the SN light-curve which are available at the time of
fitting (i.e. 2, 3 or 4 pre-maximum light epochs). The mean and
standard deviations of these distributions are shown in
Table~\ref{tab:fakesne_dist}. Fits made to the entire light-curve were
also run (i.e. including post-maximum light data); the results were
very similar to Figure~\ref{fig:synthetic-alldata-z}, with the
exception that the distributions in stretch and $\mathrm{d}m$ are
significantly tighter (see Table~\ref{tab:fakesne_dist}), as would be
expected when there is more information available to constrain these
parameters.

For these synthetic, normal SNe~Ia, the SN selection code appears
capable of recovering their parameters without any apparent bias, even
when only considering pre-maximum light data and when unknown
zeropoint uncertainties and dispersions in the SN properties are
included.  Furthermore, the estimates of redshift are good to
$\lesssim0.1$ even when only two pre-maximum light-curve epochs are
used in the fits.

However, these SN~Ia simulations do not address three important
questions.  First, do our synthetic SNe~Ia successfully represent
normal observed SNe~Ia? Is the code capable of recovering all types of
SNe~Ia without any bias against certain parts of SN~Ia parameter
space? Though our simulations generate SNe~Ia that should encompass
the full observed range of low-redshift SN~Ia properties, the
possibility that some characteristics of SNe~Ia are not represented,
or that the code may reject them as SNe~Ia, cannot be ignored.
Secondly, is our selection technique capable of distinguishing
non-SN~Ia events (e.g., core-collapse SNe) from true SN~Ia events on
early-time data (of great importance in a rolling-search)?  Thirdly,
does the choice of the cosmology used in the light-curve fit
(equation~(\ref{eq:3})) bias the SNe~Ia that are selected for
spectroscopic follow-up?  We address these questions in the next two
sections, using both the simulations described above, and using
empirical data on spectroscopically confirmed SN events discovered via
the SNLS. In Section~\ref{sec:photometric-selection-application}, we
describe the relevant aspects of the SNLS, and discuss the application
of the SN selection code to the SNLS in Section~\ref{sec:results}.

\section{Application to the SNLS}
\label{sec:photometric-selection-application}

In this section, we describe the application of the SN selection code
to real-time data obtained as part of the SNLS. The results of the
analysis, and a comparison with the simulations of
Section~\ref{sec:fakedata}, are then described in
Section~\ref{sec:results}.

\subsection{SNLS Real-time data reduction}
\label{sec:real-time-analysis}

A real-time reduction of the data are automatically performed by the
CFHT-developed Elixir data reduction system
\citep{2004PASP..116..449M}\footnote{\texttt{http://www.cfht.hawaii.edu/Instruments/Elixir/}},
which performs basic de-trending of the images, including
bias-subtraction, flat-fielding and a basic fringe subtraction in the
$i'$ and $z'$ filters. This real-time analysis is somewhat different
to that of the final Elixir reductions released through the Canadian
Astronomy Data Centre
(CADC)\footnote{\texttt{http://cadcwww.dao.nrc.ca/}}. Final processed
data are able to use all data taken during a given queue run to
construct master flat-fields and fringe frames, which is not possible
in real-time where library versions of these calibration frames from
the previous queue run are used. In particular, flat-fielding and, in
the case of $z'$ data, fringe subtraction are both inferior in
real-time data.

The Elixir-reduced data are then processed through two independent
search pipelines, written by members of the SNLS collaboration in
Canada and France, from which combined candidate lists are generated.
The two pipelines typically produce candidate lists which agree at the
90\% level down to $i'\simeq24.0$. In this paper, we refer exclusively
to the Canadian pipeline, though the details of the French pipeline
are similar in most respects. Full details of these reductions will be
presented in forth-coming papers; here we present a brief overview. We
perform a photometric alignment to secondary standards within each
field using a multiplicative scaling factor, and astrometrically align
the data in different filters using custom-built astrometric reference
catalogs.  Finally, the frames are re-sampled to a common pixel
coordinate system, and a median stack in each filter is generated,
rejecting cosmic-ray events and other chip defects.

The point-spread functions (PSFs) of the image stack and the deep
reference image constructed from observations from the previous year
are matched using a variable kernel technique, and a subtraction image
generated containing only sources which have varied since the
reference epoch.  This subtraction image is searched for SN candidates
using automated techniques, with all likely candidates screened by a
human eye to ensure obviously non-stellar sources are not included on
the potential candidate follow-up list.

PSF-fit and aperture photometry for all new and previously detected
sources is measured and the flux information is stored in a
publicly-available database of all SN
candidates\footnote{\texttt{http://legacy.astro.utoronto.ca/}}.  New
candidates are automatically cross-correlated with the database of
existing variable sources (to identify previously known AGN and
variable stars), and back-tracked through previous epochs in order to
construct as much information as possible as to their nature.
Additionally, each candidate is given a provisional category of either
likely supernovae (SN), likely AGN (if the detection is precisely
centered and appears intermittently over long periods of time), likely
variable star (if the candidate host appears stellar and PSF-like), or
likely moving object (if the candidate moves between epochs). Where
any reasonable doubt exists, the SN classification is retained.  The
result is a database of flux information for every variable candidate
detected in the SNLS.

Spectroscopic follow-up for the SNLS is currently provided by
dedicated follow-up programs on the European Southern Observatory Very
Large Telescope (VLT; 60 hours/period), Gemini North and South (60
hours/semester) and Keck-I/II telescopes (4 nights in the ``A''
semester of each year), as well as other programs at Keck-I making
detailed studies of select SNe (4-8 nights a year). During the
time-frame covered by this paper, the VLT observed 61 candidates (Basa
et al., 2005 in prep.), Gemini observed 40 candidates
\citep{2005astro.ph..9195H} and Keck observed 21 candidates, 7 as part of
routine follow-up, and 14 for a detailed study program (Ellis et al.,
in prep.) Some candidates were observed on more than one occasion.
Full details of the reductions and SN typing methods employed can be
found in the papers referenced above. Typically, redshifts are
determined from host galaxy spectral emission or absorption features,
or via a $\chi^2$ template matching technique to a large library of
local SN spectra of all types \citep{2005astro.ph..9195H}.

\subsection{Analysis}
\label{sec:analysis}

The photometry and candidates that we use in this analysis are taken
from the Canadian real-time analysis corresponding to the period May
2004 to January 2005 -- or approximately 3 ``field-seasons''. Over
this period, the SN selection code was used to assist in the selection
of spectroscopic follow-up candidates. 440 potential SN candidates
were located by our real-time analysis software (excluding likely AGN
and variable stars), of which 121 were followed spectroscopically.  Of
these, 70 have since been identified as SNe~Ia or probable SNe~Ia
(denoted SN~Ia* hereafter), 6 as being consistent with an SN~I, 11 as
SNe~II, 4 as SN~Ib/c.  30 candidates remain unidentified or are non-SN
events.  During this period, in order to test the SN selection code,
in some months we deliberately followed events that were predicted
\textit{not} to be SNe~Ia to ensure we were not biasing against
particular classes of SN~Ia events.

For this analysis, we run the SN selection code on all SN candidates
discovered during this period (fixing \ebmvmw\ as appropriate for each
candidate), firstly fitting on the entire real-time light-curve, and
secondly fitting just the data that were available at the time the
follow-up decision was made. This exact date was not accurately
recorded or easily reconstructed for all events; we approximate this
by cutting-off the real-time light-curves at 2 days before maximum
light in the SN rest-frame.  Clearly, this cut restricts us to only
consider candidates with at least 2 epochs prior to 2 days before
maximum light, $\sim95$\% of the total SN~Ia sample; furthermore, the
approximation may lead to some fits using more data (and some less)
than was actually used at the time the follow-up decision was made.

\section{Results}
\label{sec:results}

\subsection{Discriminating core-collapse SNe from SNe~Ia}
\label{sec:discrimating-cc}

We first examine the non-SN~Ia events located by the SNLS. One important
requirement of the SN selection technique is to not only return the
most likely parameters describing a SN based on the fitting the
real-time light-curve to a normal SN~Ia model, but also to return some
guide as to the likelihood that the target is indeed a normal SN~Ia.
As fits to non-SNe~Ia should be of lower quality, we examine the
$\chi^2$ of the light-curve fits to examine whether it can be used as
a reliable indicator.

The most common form of contamination in our sample comes from
core-collapse SNe. While AGN and variable stars are obviously present,
they are usually easily screened out as they show significant
variation on the time-scale of a year, and are usually already present
in our database of variable objects. Core collapse SNe -- the various
sub-classes of SNe~II and SNe~Ib/c -- are heterogeneous; the range in
peak magnitudes \citep{2002AJ....123..745R}, spectra
\citep{2001AJ....121.1648M}, and colors can be considerable.  The
general trend however is that for a core-collapse SN and a SN~Ia
located at the same redshift, the core collapse SN will be fainter
than the SN~Ia. Typical SNe~II are 1-2.5\,mag fainter at peak than
SNe~Ia, and typical SNe~Ib/c around 1.5\,mag fainter \citep[see table
1 in ][]{2002AJ....123..745R}, though there are examples of more
luminous or peculiar events in the literature
\citep[e.g.][]{2000ApJ...533..320G,2000ApJ...534L..57T}.

\begin{figure} 
\plotone{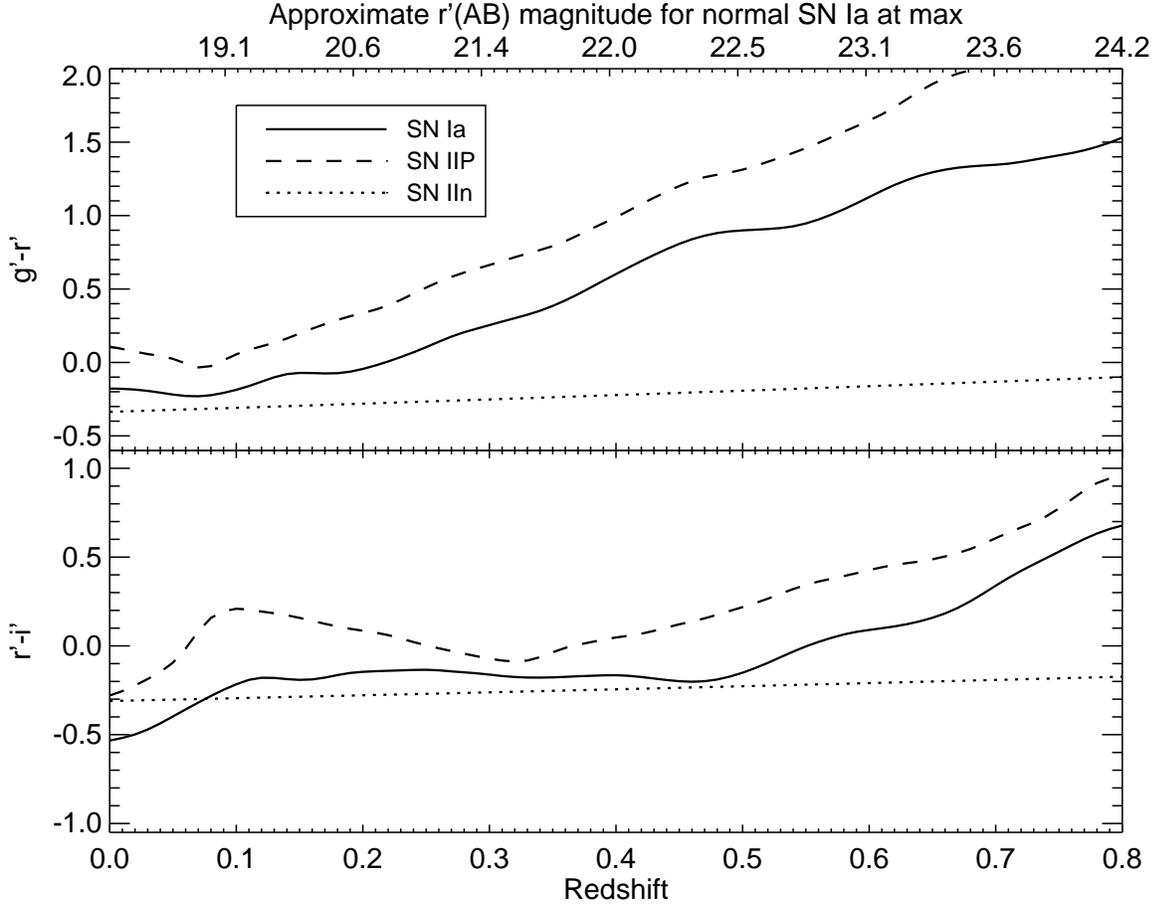}
\caption{
  The observed $g'-r'$ and $r'-i'$ colors of SNe~Ia compared to some
  SNe~II models as a function of redshift. As SNe~Ia typically appear
  1-2\,mag brighter than SNe~II located at the same redshift, the
  strong evolution in SN~Ia $g'-r'$ color with redshift provides a
  strong discriminant against core-collapse SNe, which appear too blue
  for a redshift based on a SN~Ia fit.
  \label{fig:colz}}
\end{figure}

\begin{figure} 
\plotone{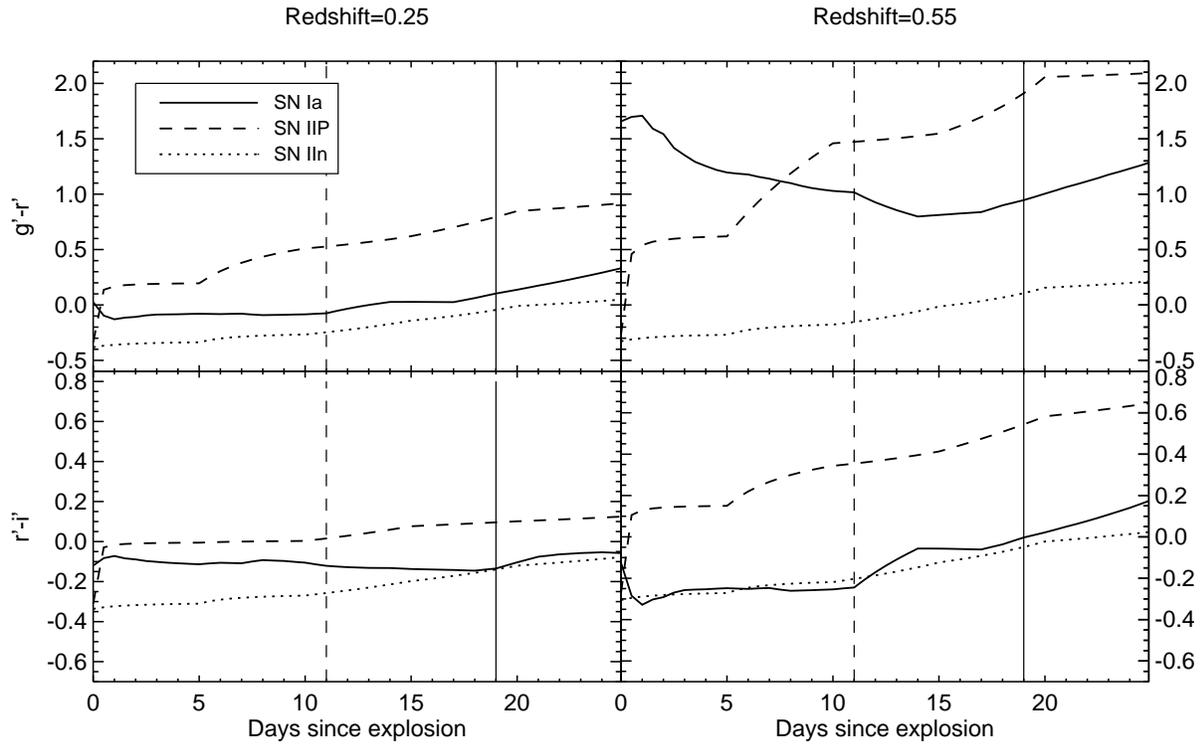}
\caption{
  The observed $g'-r'$ and $r'-i'$ colors of SNe~Ia compared to some
  SNe~II models as a function of rest-frame phase of the SN. Two
  redshifts are shown; $z=0.25$ and $z=0.55$. Note that SNe~II and
  SNe~Ia peak at different times after explosion; these are indicated
  with different vertical lines in the figure.
  \label{fig:colphase}}
\end{figure}

Our SN~Ia selection technique uses both this peak magnitude
information and the color of the observed light-curve, fitting a given
candidate for redshift, phase, stretch, extinction, and dispersion in
the peak magnitude under the assumption that the object is a SN~Ia in
an assumed cosmology. This places a constraint on the redshift that
can be fit (as SNe~Ia are calibrated candles and have similar
intrinsic luminosities), and hence restricts the range of observed
colors that are consistent with a SN~Ia as a function of SN
brightness.  We demonstrate this in Fig.~\ref{fig:colz}, where we show
the redshift evolution in the color of a SN~Ia and two example models
for a SN~II (in this case an SN~IIP and an SN~IIn), and
Fig.~\ref{fig:colphase}, where we show the evolution of SN colors as a
function of phase. The SN~Ia template is that discussed in
Section~\ref{sec:SN-model} and the core-collapse templates are
described in Section~\ref{sec:simulating-snls}.

Consider a SN~IIP at maximum light located at $z=0.25$ with
$g'-r'\sim0$ and $r'\sim23$.  Though the color is also consistent with
a maximum-light SN~Ia at the same redshift (Fig.~\ref{fig:colz}), the
peak magnitude of the SN~IIP is around 2 magnitudes fainter than a
SN~Ia would be ($r'\sim21$). This $r'\sim23$ peak magnitude
\textit{is} consistent with a SN~Ia at $z\simeq0.55$; however, at this
redshift the $g'-r'$ color of a normal SN~Ia is $\sim$1\,mag redder.
Hence, SNe~II (and other core-collapse events) can potentially be
screened out by searching for objects that appear too blue in $g'-r'$
for the fitted redshift when compared to the SN~Ia model, or more
precisely, rejecting objects that have a large $\chi^2$ per degree of
freedom (d.o.f) in the $g'$-band from the SN~Ia fit. A measurement of
color at exactly maximum light is not required; the same trend is
apparent up to about 20 days after explosion
(Fig.~\ref{fig:colphase}).

\begin{figure} 
\plottwo{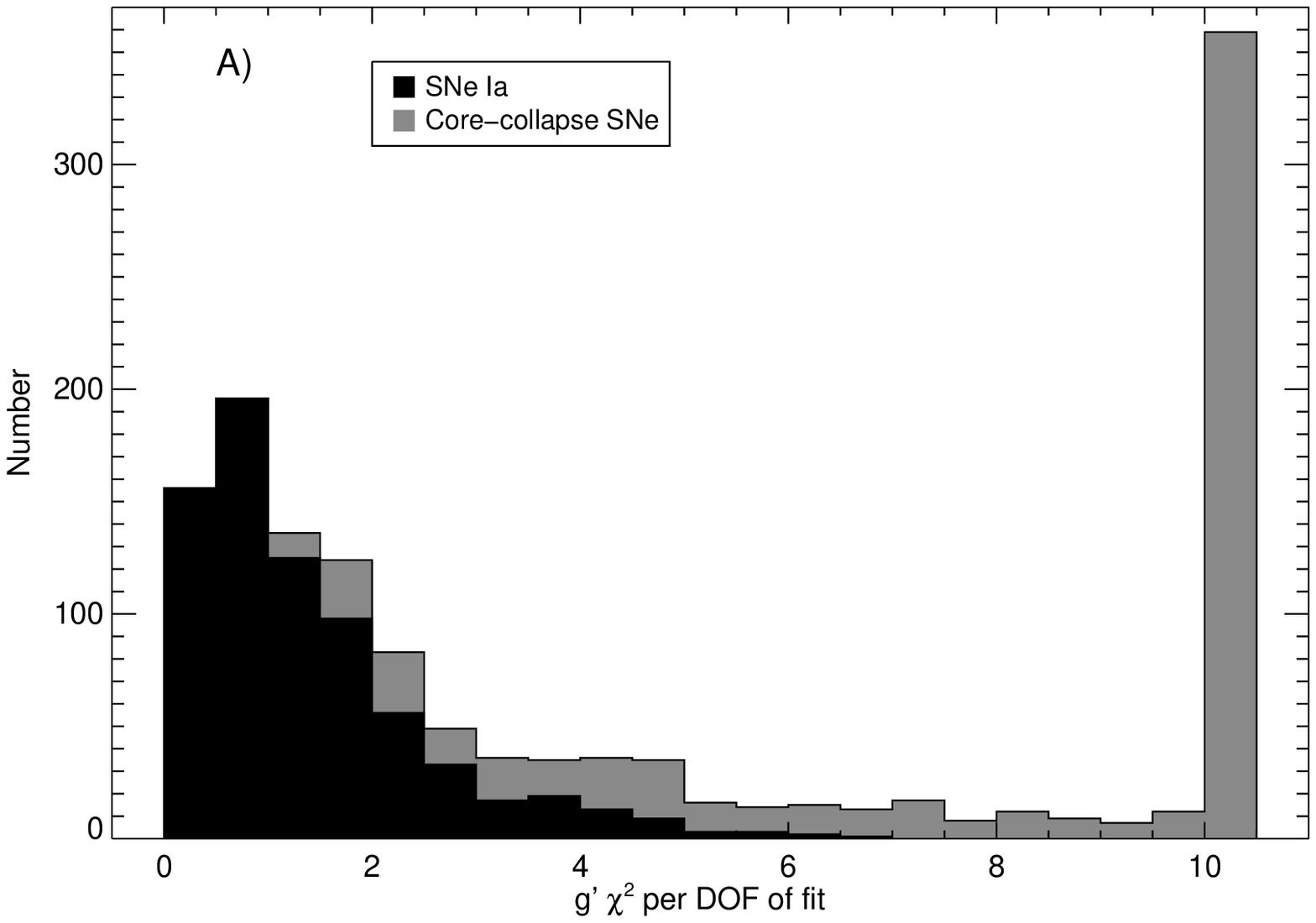}{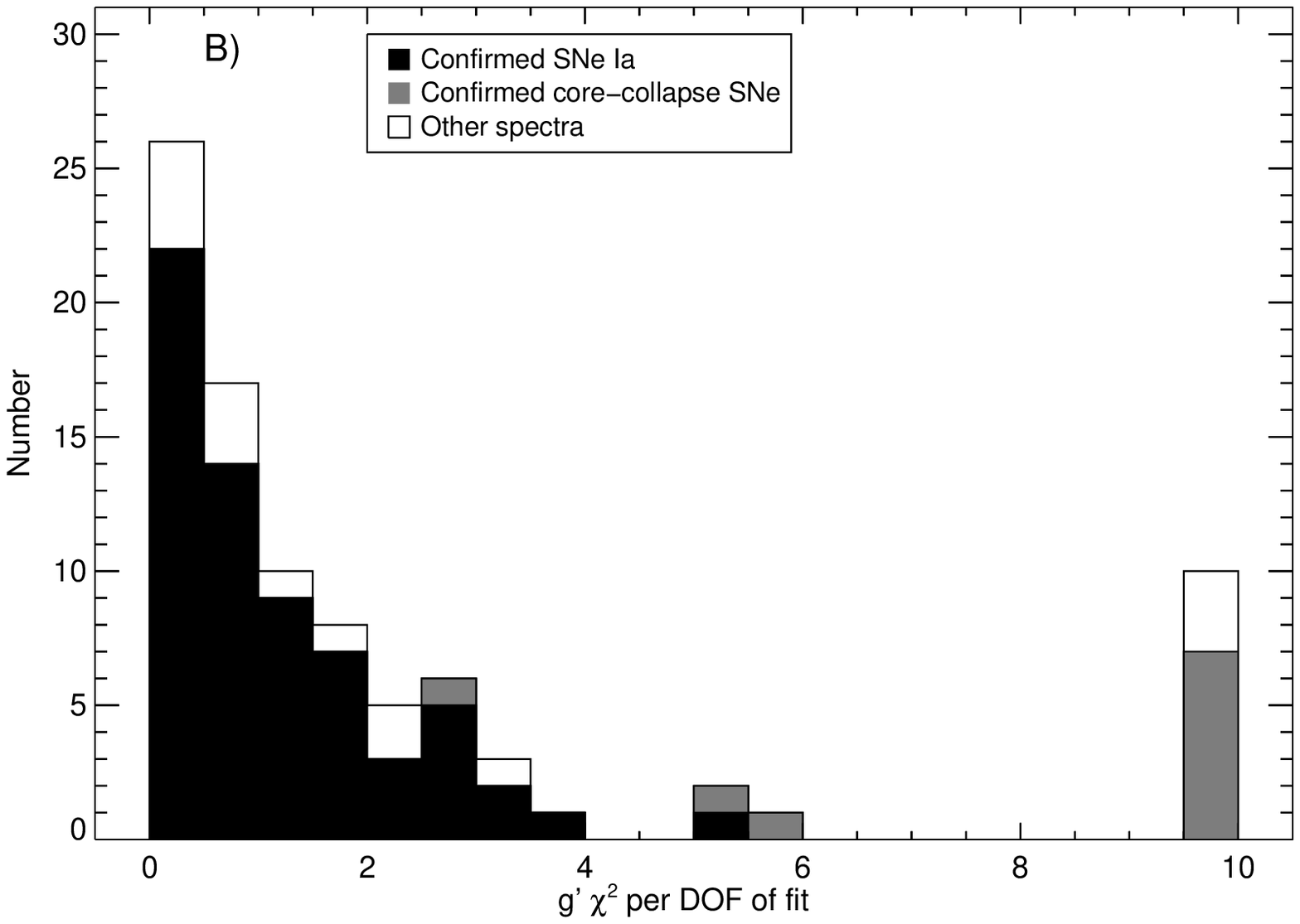}
\caption{
  The distribution of $g'$ $\chi^2$ per d.o.f. for the fits of all SN
  candidates for the simulations (Panel a, left;
  Section~\ref{sec:fakedata}) and for real SNe discovered by SNLS
  (Panel b, right). In panel a) `Core collapse SNe' represent all the
  various sub-types of core-collapse SNe as described in the text. In
  panel b), `Confirmed SNe~Ia' are those SNe identified as either `Ia'
  or `Ia*' from their spectra, `Confirmed core collapse SNe' are those
  identified as any variant of a SN~II, `Other spectra' includes any
  other object observed spectroscopically regardless of the type of
  object.
  \label{fig:gchi2-hist}}
\end{figure}

\begin{figure} 

  \includegraphics[width=2.1in]{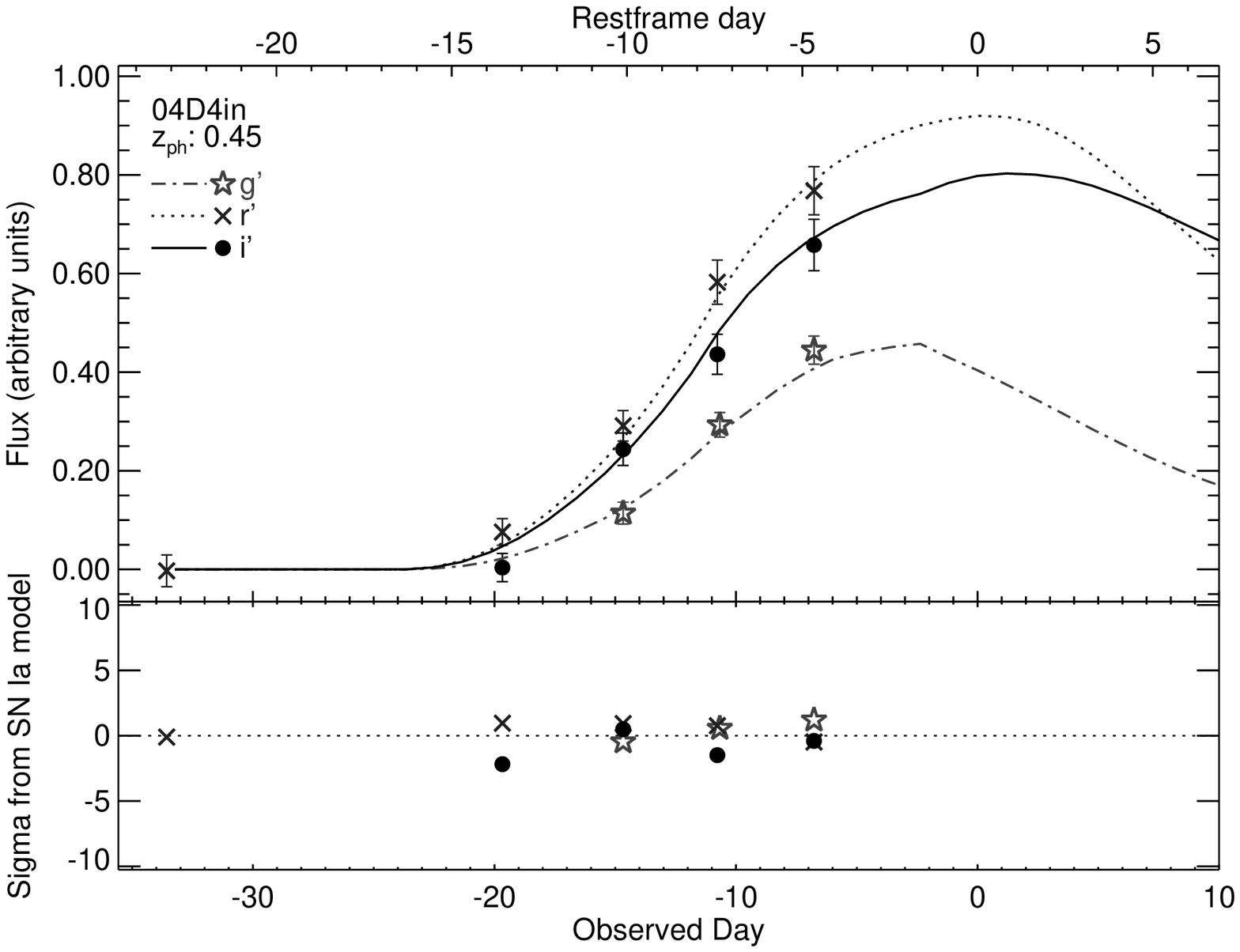}
  \includegraphics[width=2.1in]{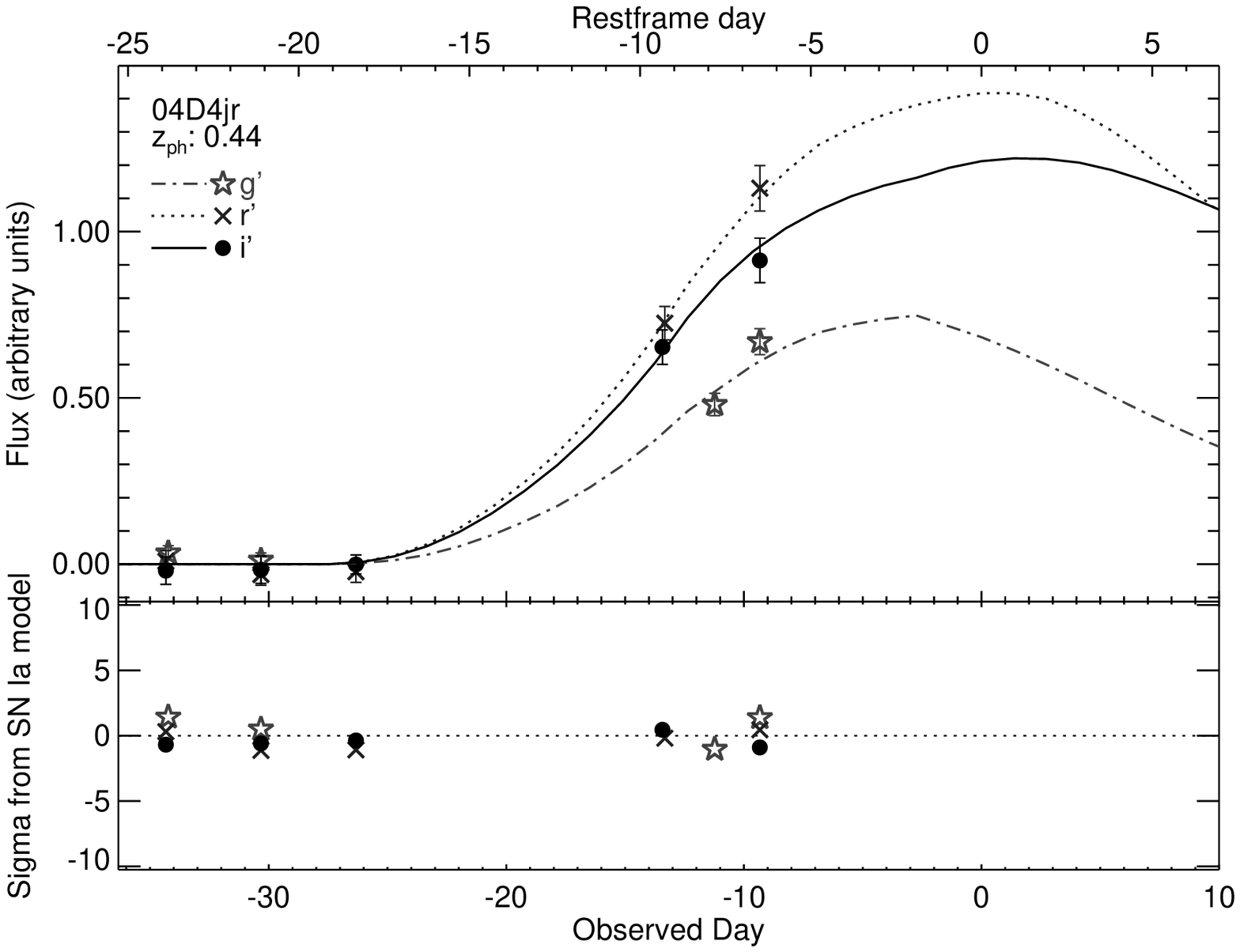}
  \includegraphics[width=2.1in]{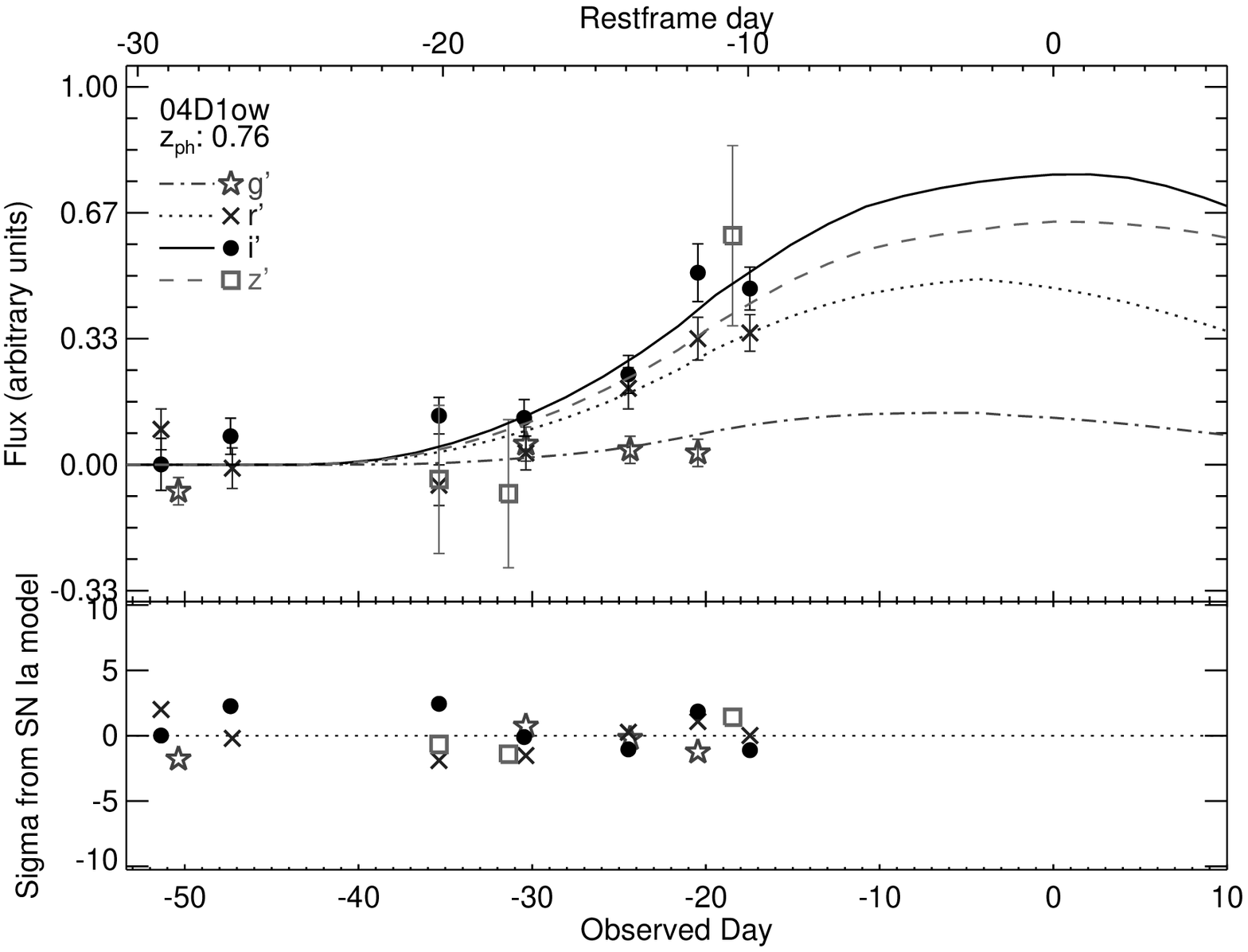}

  \includegraphics[width=2.1in]{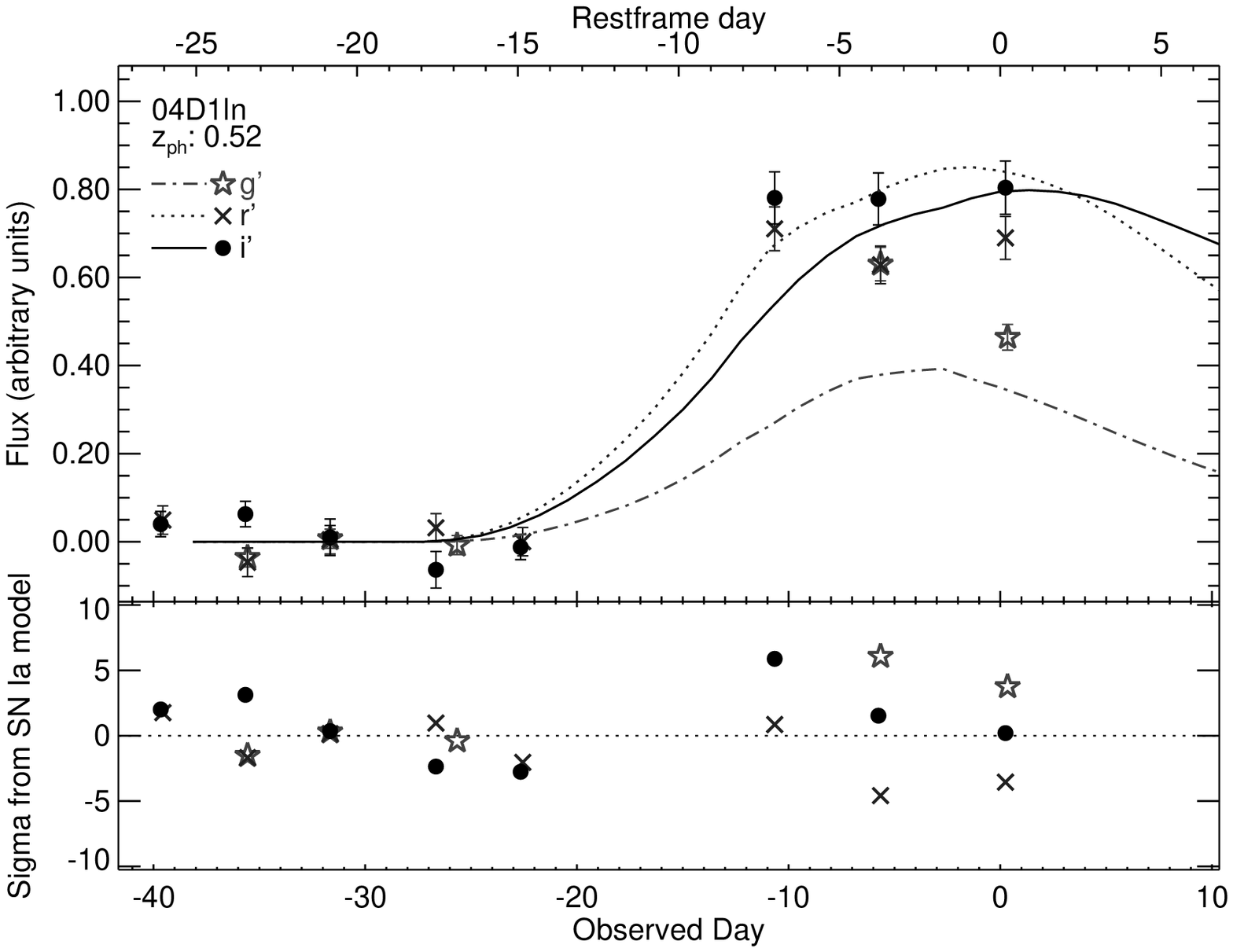}
  \includegraphics[width=2.1in]{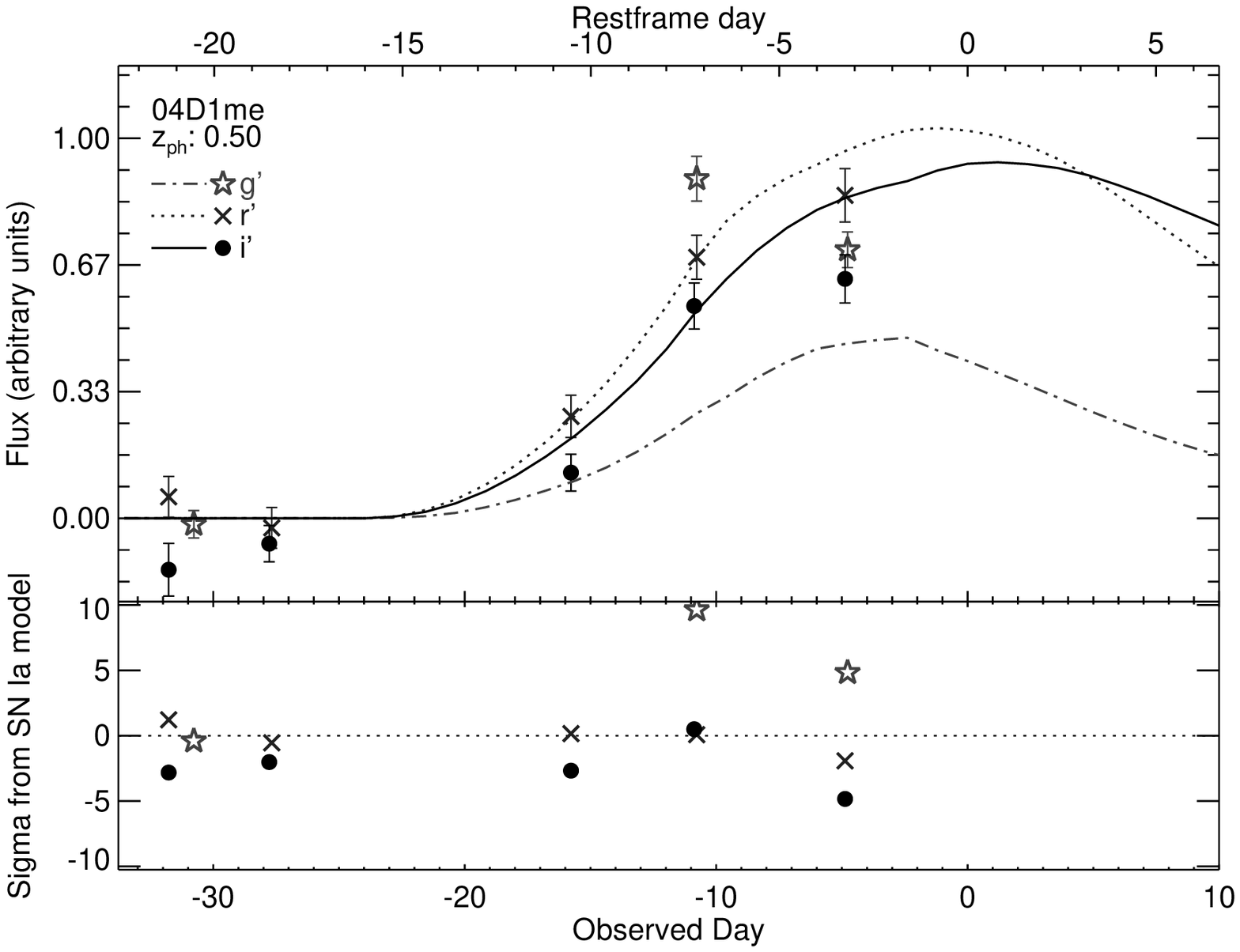}
  \includegraphics[width=2.1in]{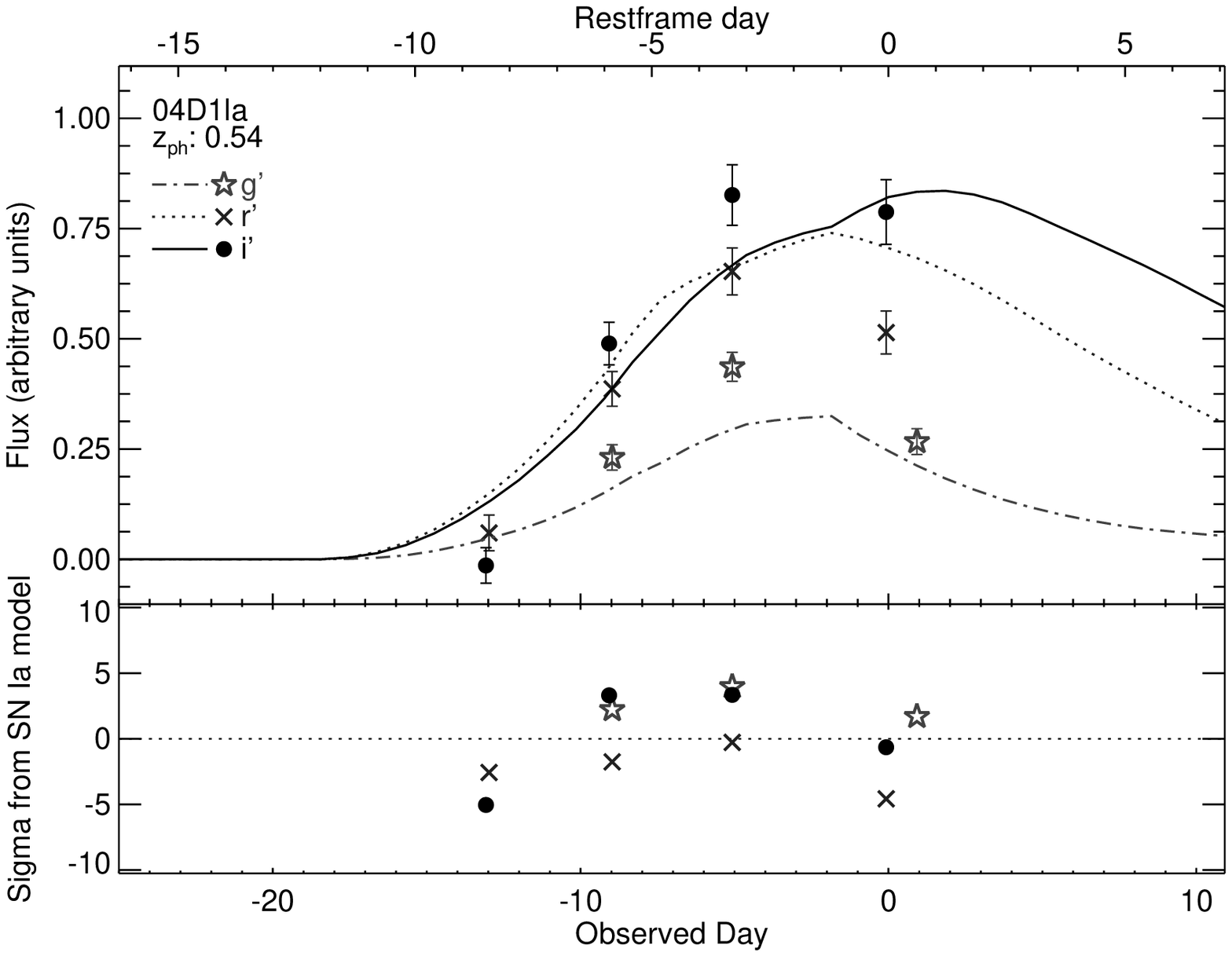}

\caption{
  Example light-curve fits to early real-time photometry. In all cases
  the light-curves are fit to a SN~Ia model, with the phase, redshift
  and stretch free parameters in the fit as described in the text. The
  top row shows fits to known SNe~Ia; from left to right 04D4in
  ($z=0.52$), 04D4jr ($z=0.48$), and 04D1ow ($z=0.92$). The bottom row
  shows fits to known core-collapse SNe; from left to right: 04D1ln
  (SN~IIP; $z=0.21$), 04D1me (SN~IIP; $z=0.30$), and 04D1la (SN~Ib/c;
  $z=0.32$). In all cases, the deviations from the SN~Ia model are
  significantly larger for the core-collapse SNe, allowing an
  efficient identification of such objects based on the $g'$ $\chi^2$.
  \label{fig:example_lcs}}
\end{figure}

The distributions of $\chi^2$ in the $g'$-band from our pre-maximum
light fits to the simulated and real SNLS candidates are shown in
Fig.~\ref{fig:gchi2-hist}. Example light-curve fits to both real
SNe~Ia and real core collapse SNe are shown in
Fig.~\ref{fig:example_lcs}. As expected, the SNe~Ia and core collapse
SNe have quite different distributions, both for the simulated survey
data, and for the real survey data.  SNe~Ia typically have a small
$\chi^2$; the median value is 0.85 (1.01 for the simulated data), and
90\% of SNe~Ia have $\chi^2<2.80$ ($\chi^2<2.91$ in the simulations).
Only 1\% (0.9\% in simulated data) have $\chi^2>5$. 

The $\chi^2$ for core collapse SNe is typically larger, with 10 of 11
events having a $\chi^2>5$; all SNe~II have $\chi^2>5$. The numbers for
the fits to the simulated SNe are similar; 80\% of simulated
core-collapse SNe have a $\chi^2>5$. Thus, the $\chi^2$ in the
observed $g'$ filter seems an efficient discriminant with which to
remove the majority of core collapse SNe from spectroscopic follow-up
samples.

\subsection{Redshift and phase precision}
\label{sec:rta-redshiftprecision}

The accuracy to which redshift and phase can be estimated are
important quantities; the first allows an optimal spectrograph setting
to be used, and the second ensures that SNe are observed at a phase
when the different sub-types show the most diversity and with the
greatest contrast over their host galaxy. Here we investigate how well
the predicted redshift and phase of the SNe~Ia agree with actual
values.

\begin{figure} 
\plotone{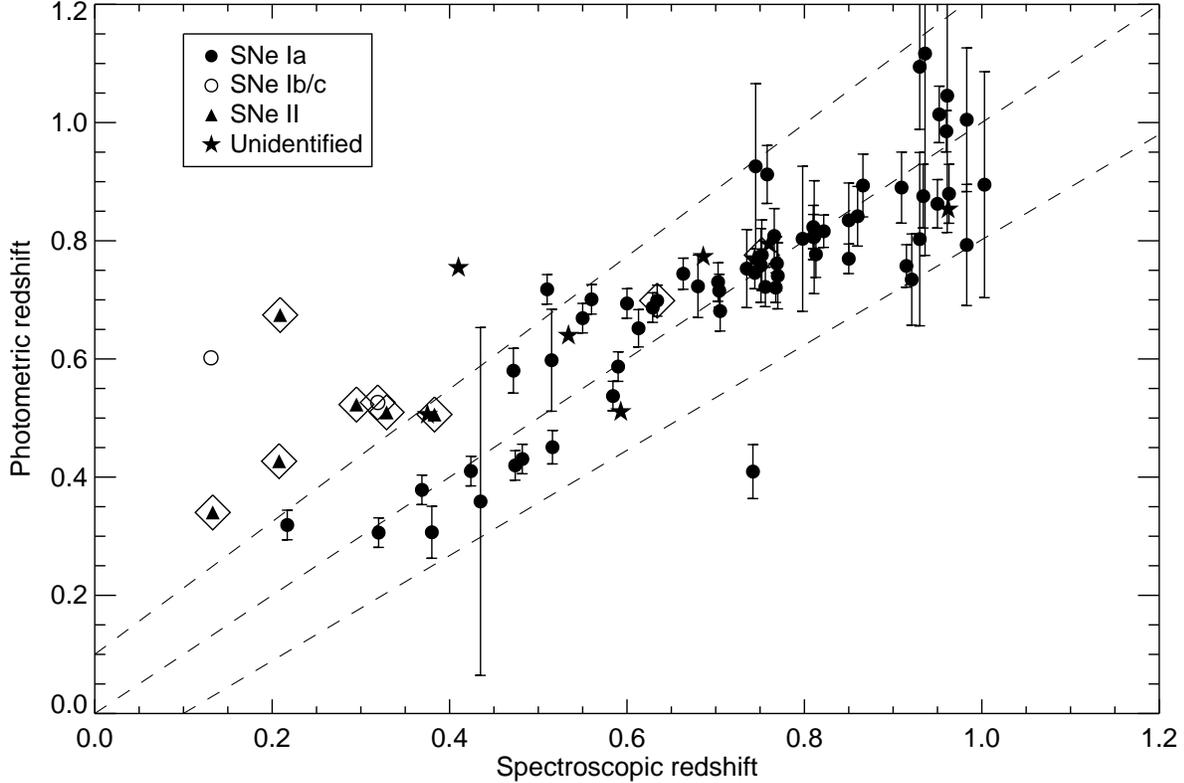}
\caption{
  The comparison between \zphot\ and \zspec\ for the candidates
  followed as part of the SNLS. The photometric redshifts are from the
  SN and are based on pre-maximum-light data only. The SN are coded
  according to the type of the event, and those events that would have
  been removed from our follow-up sample based \textit{purely} on the
  $g'$ $\chi^2$ (per d.o.f.) from the SN~Ia fit are boxed with a
  diamond.  The two rejected SNe~Ia have only poor quality pre-maximum
  light data; in reality more data was utilized for these candidates
  before a follow-up decision was made. The \zphot\ errors are those
  reported by the least-squares minimizer. The over-plotted dashed
  lines show a 1:1 correspondence between \zphot\ and \zspec\ as well
  as $\pm$10\% in $(1+z)$. The assumed cosmology is $\omatter=0.25$,
  $\olambda=0.75$, and the SNe~Ia include both `Ia' or `Ia*' events.
  \label{fig:speczphotozLCDM}}
\end{figure}

The photometric-redshift (\zphot) versus spectroscopic-redshift
(\zspec) comparison for all spectroscopically observed objects where a
\zspec\ could be determined is shown in
Fig.~\ref{fig:speczphotozLCDM}. The plot is based on fits to just the
data that was available at the time the follow-up decision was made
(taken as pre-maximum light data).

\begin{figure} 
\plotone{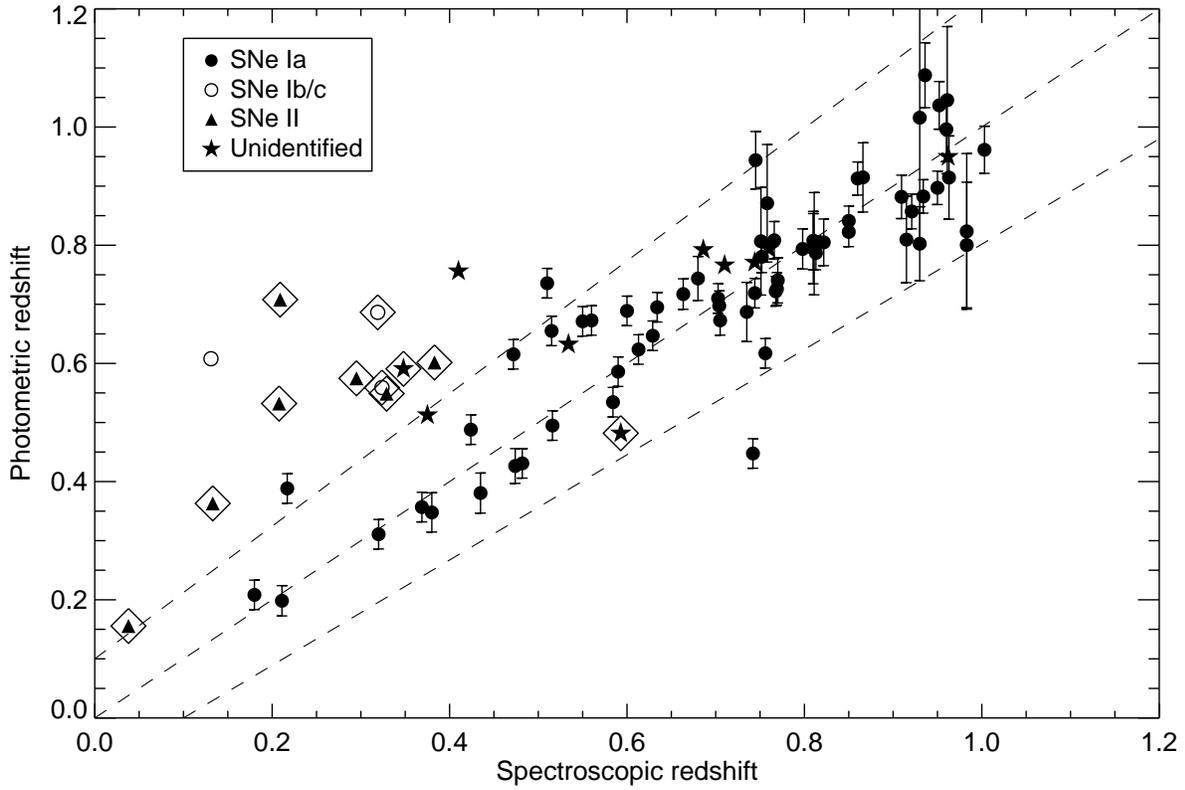}
\caption{
  As Fig.~\ref{fig:speczphotozLCDM}, but using the entire light-curve
  in the fit rather than just the pre-maximum light data. Note that
  more SNe appear on this plot than on Fig.~\ref{fig:speczphotozLCDM},
  as this plot includes SNe for which there is no pre-maximum light
  data.
  \label{fig:speczphotozLCDM_all}}
\end{figure}

The agreement between \zphot\ and \zspec\ for SNe~Ia is remarkably
good. The median $\left|\deltaz\right|/(1+\zspec)$ is 0.031, with 90\%
of the SNe~Ia having $\left|\deltaz\right|/(1+\zspec)<0.08$. The
agreement for other SN types is, as expected, much poorer, as the
differing colors, light-curves shapes and brightnesses of these SN
leads to an incorrect redshift estimate.  However, most of these core
collapse candidates are rejected by the code as being SNe~Ia due to
the $\chi^2$ of the fits; such rejected SNe are marked on the figure.
As a comparison, Fig.~\ref{fig:speczphotozLCDM_all} shows the \zphot\ 
versus \zspec\ for fits based on the entire SN light-curve; here the
median $\left|\deltaz\right|/(1+\zspec)$ is 0.025. Note that the
apparent high accuracy of this technique does not mean that
spectroscopic redshifts are not needed for cosmological analysis;
clearly as a cosmology is assumed in order to derive the fitted
redshift, this redshift could not then be used to re-derive
cosmological constraints. Furthermore, as the nature of the events are
not known to be SNe~Ia \textit{a priori}, and the assumed cosmology is
a very useful prior when identifying the core collapse SNe, the
technique could not be used to derive cosmological constraints
\citep[as can be done in some low-redshift SN~Ia samples,
e.g.][]{2004ApJ...613L..21B} without a significant improvement in our
knowledge of the color evolution and light-curve morphology of core
collapse SNe.

\begin{figure} 
\plotone{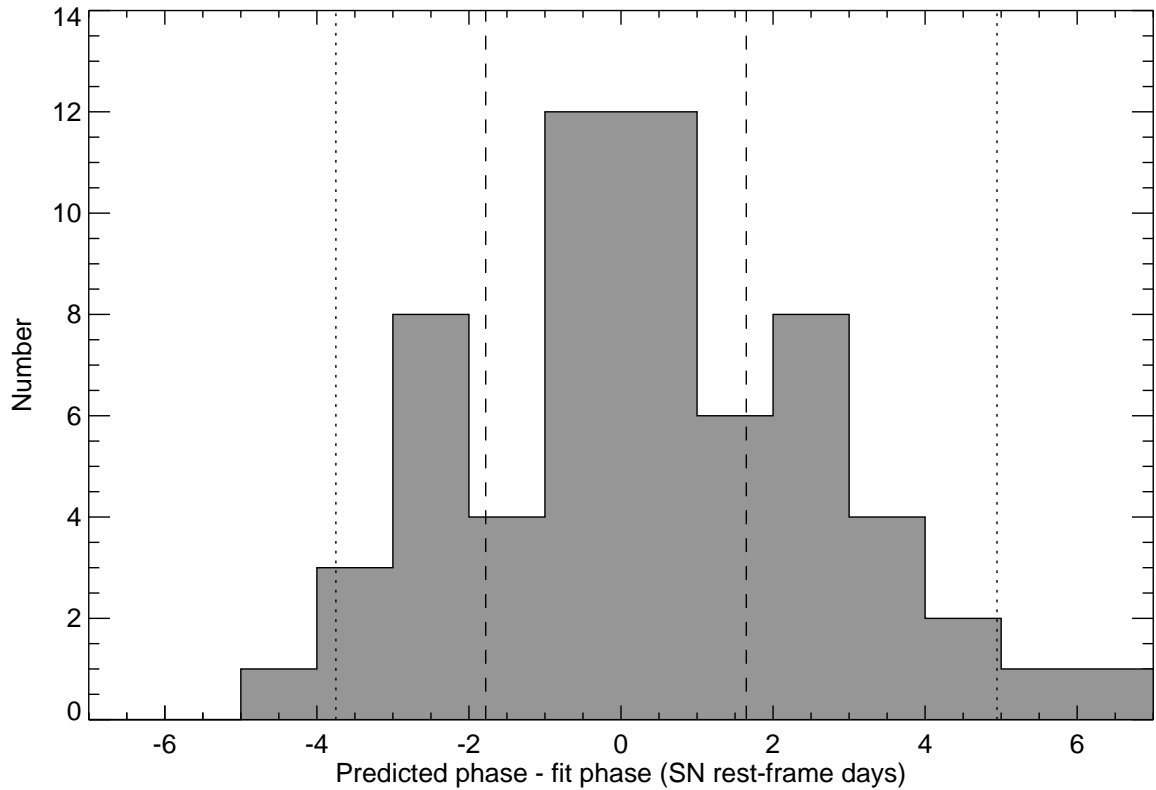}
\caption{
  The distribution of the predicted time of maximum light minus the
  actual time of maximum light in the SN rest-frame. The ``predicted
  time of maximum light'' is based only on pre-maximum light
  light-curve data i.e. the data available at the time of
  spectroscopic follow-up.  The ``actual time of maximum light'' is
  based on the entire SN light-curve. The vertical lines denote the
  region containing 50\% (dashed) and 90\% (dotted) of all candidates.
  Only confirmed SNe~Ia (or SNe~Ia*) with spectroscopic redshifts are
  shown. The uncertainty on the time of maximum light from fits based
  on the entire light-curve is $\sim 0.5-1.0$\,days.
  \label{fig:phasecompare}}
\end{figure}

The distribution of the predicted date of maximum light versus the
date of maximum light based on the entire light-curve is shown in
Fig.~\ref{fig:phasecompare} (as before, predicted values are only
based on candidate light-curves up to 2 rest-frame days before maximum
light). The agreement is good, with 50\% of the predicted times of
maximum light lying within $\pm1.75$ rest-frame days of the actual date
of maximum light, and  90\% within $\pm\sim4$ days.

\subsection{Effect of assumed cosmology}
\label{sec:effect-of-cosmology}

In this section, we investigate any possible bias in the population of
objects that would be selected for spectroscopic follow-up based on
the ``assumed'' cosmology that is used in the light-curve fit. We
repeat the SN selection fits on the real survey data using a different
cosmological model, and compare the identity of the rejected
candidates with those rejected when using the standard cosmology.

Subtle changes in the cosmology, for example varying the value of the
equation of state parameter $w$, have no effect on the nature of the
candidates that are rejected. For example, the candidates that are
rejected (or selected) with $w=-1$ are also rejected (or selected)
with $w=0.8$ or $w=-1.2$, as the differences in the
apparent-magnitude/redshift relation are very small. A more drastic
test is to compare to a cosmology that makes very different
predictions in the apparent magnitude of an object as a function of
redshift. We compare to an Einstein-de-Sitter (EdS) cosmology
($\omatter=1.0$, $\olambda=0.0$); the apparent magnitude of a SN~Ia
differs by $\simeq0.5$\,mag at $z=0.6$. Any bias in the objects that
are selected for follow-up by assuming an accelerating universe
cosmology should be evident after these fits.

\begin{figure} 
\plotone{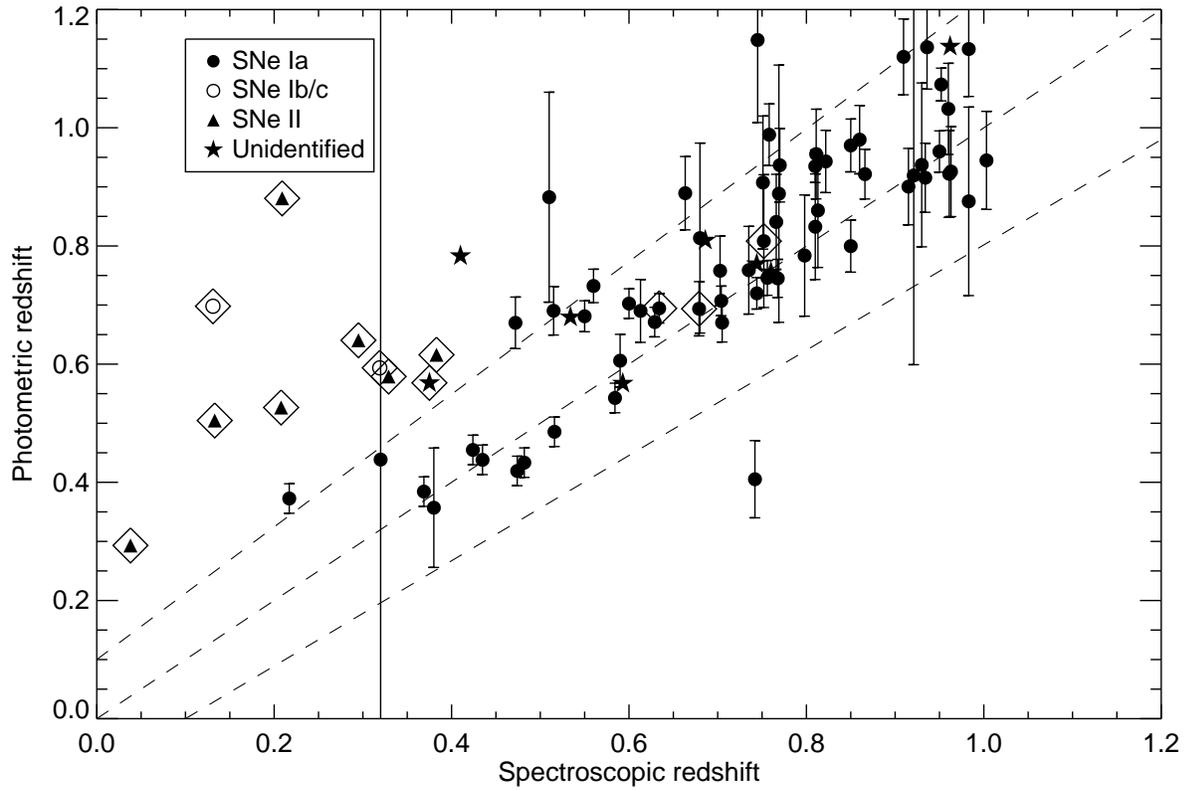}
\caption{
  As Fig.~\ref{fig:speczphotozLCDM}, but assuming an $\omatter=1.0$,
  $\olambda=0.0$ cosmology. Changing the assumed cosmology does not
  alter the SNe that are rejected as being SNe~Ia by the fitting code,
  though the agreement between \zspec\ and \zphot\ is poorer.
  \label{fig:speczphotozEdS}}
\end{figure}

In Fig.~\ref{fig:speczphotozEdS} we show the same $z_{\mathrm{phot}}$
versus $z_{\mathrm{spec}}$ comparison as in
Fig.~\ref{fig:speczphotozLCDM}, but for the EdS Universe. There is a
clear offset apparent in these plots; in an EdS Universe, the \zphot\ 
is found to be systematically larger than the \zspec, and the size of
this effect increases at higher redshift. This is simple to
understand; in a EdS universe, objects appear brighter than in a
$\Lambda$-cosmology at the same redshift. If we live in a
$\Lambda$-dominated Universe, the SN selection code will over-estimate
the redshift (i.e. will try to make the objects fainter) if it assumes
an EdS cosmology.

The objects that are rejected based on the $g'$ $\chi^2$ are the same
when assuming an EdS Universe as when assuming a $\Lambda$-cosmology
i.e. the assumed cosmology does not place a strong constraint on the
objects that are selected for spectroscopic follow-up. However,
assuming a $\Lambda$-cosmology does improve our estimates of candidate
redshifts and hence stretches and observer-frame ages, which is
essential to ensure that candidates are observed at the optimal time
\citep{2005astro.ph..9195H}. The assumed cosmology does not provide the key
discriminant in rejecting non-SN~Ia events, but does enable a more
efficient follow-up observation to be performed.

\subsection{Implications for cosmological measurements}
\label{sec:cosm-impl-sn}

We now examine the SNe~Ia that are rejected in any cosmology, i.e.
SNe~Ia that have sufficiently discrepant stretches, luminosities, or
colors when compared to the standard template that is used in the
fits, or SNe where the selection code simply fails to find the correct
solution. If similar sub-types of SNe~Ia were consistently rejected,
this could introduce a bias in the types of SNe~Ia that were sent for
follow-up, with a possible consequent bias on the cosmological results
that would be derived from the resulting spectroscopic sample.
Alternatively, if the occasional rejections of a SN~Ia were random,
and related to the phasing of the observations or the poor quality of
a given epoch of real-time data, then the impact on the derived
cosmological parameters would be very small.

We investigate this effect using a simulated dataset similar to that
in Section~\ref{sec:fakedata}. We generate 700 high-redshift
($0.15<z<0.9$) SNe~Ia (the SNLS end-of-survey goal), and combine with
300 local SNe~Ia ($0.01<z<0.12$), a sample size equivalent to that
likely to be available at the end of the SNLS from projects such as
the Carnegie Supernova Project \citep{2005ASPC..339...50F} and the
Nearby Supernova Factory \citep{2002SPIE.4836...61A}. The 700
high-redshift SNe are fit using the SN selection code, the pre-maximum
light data fit, and those SNe which are formally rejected are noted.
Typically 1-2\% are formally rejected by the code, usually because of
a lack of data before maximum light (due to weather in our
simulations) rather than an intrinsic property of the SN. We then
perform a full cosmological fit of $w$ and \omatter, firstly on the
full 700+300 SN~Ia sample, and then on the same sample less the few
high-redshift SNe that were rejected by the SN selection code based on
the pre-maximum light fits. The cosmological fits follow a similar
method as that used in \citet{2005astro.ph.10447A}, and produce
confidence contours in $w$ and \omatter. In all cases, the input
cosmological values were $\omatter=0.3$ and $w=-1$, and a flat
Universe was assumed.

\begin{figure} 
\plotone{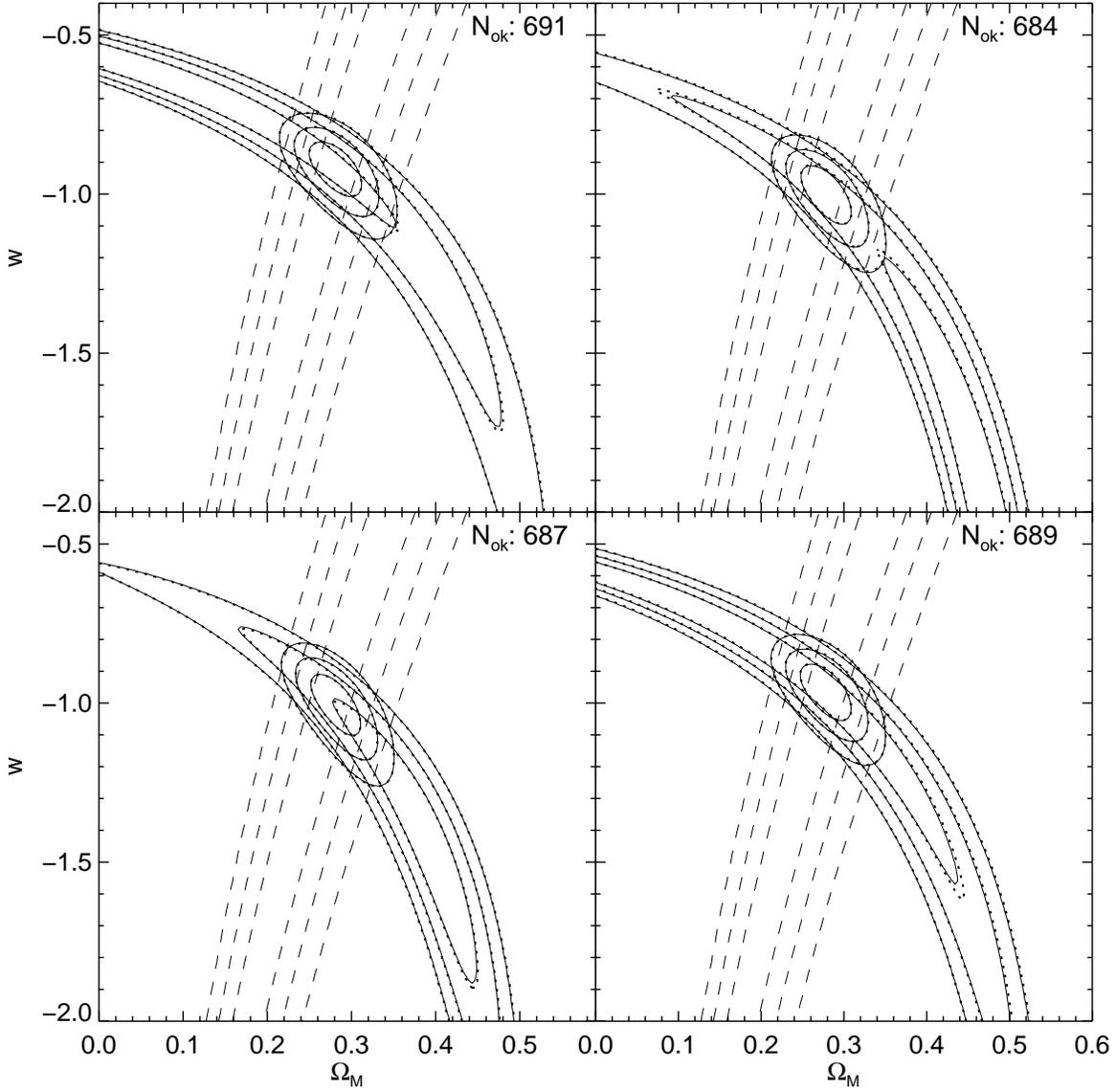}
\caption{
  The cosmological fits in $w$ and \omatter\ to four simulated
  SNLS-like SN samples with and without the few SNe~Ia formally
  rejected by the SN selection code. Each panel represents a different
  SNLS-like survey; the differing positions of the contours is
  expected and arises from random variation from simulation to
  simulation. In each panel, the solid contours indicate the full
  sample of 700 SNLS SNe plus 300 low-redshift SNe, the dotted lines
  are the same sample minus those high-redshift SNe rejected by the SN
  selection code. In each case, the confidence contours with and
  without the baryon acoustic oscillations constraint \citep[dashed
  line;][]{2005astro.ph..1171E} are shown.  The number in each panel
  indicates the number of high-redshift SNe not rejected by the
  selection code.
  \label{fig:cosmo_fits}}
\end{figure}

As each survey simulation generates different SNe, we repeat the full
simulation four times; the confidence contours for the recovered
cosmological parameters in \omatter\ and $w$ are shown in
Fig.~\ref{fig:cosmo_fits}. Constraints from the baryon acoustic
oscillations (BAO) are also included \citep{2005astro.ph..1171E}. The
differences between the cosmological fits for the different samples
are very small. In all cases, the difference in the derived value of
$w$ is less than 0.001 ($\sim$0.02$\sigma$) when including the BAO
constraints, and less than 0.015 ($\sim$0.05$\sigma$) when the BAO
constraints are not used. We therefore conclude that the SN selection
method adds no significant bias to the eventual cosmological results
of the SNLS.

\subsection{Potential SN~Ia targets not followed spectroscopically}

An alternative, more empirical, method of how successful the SN
selection code is at locating candidates for follow-up in real-time,
is to examine the fraction of good SN~Ia candidates rejected by the
code. The impact on the determination of the cosmological parameters
of these SNe was shown to be negligible in the previous section;
however, this is also important for future studies of the SN~Ia that
rely on near-complete spectroscopic samples of SNe~Ia being available
(e.g., SN rates).  We assess this by examining the distribution of the
fitted SN stretches of our candidates once their light-curves are
complete.  Virtually all normal SN~Ia stretches lie in the range 0.7
to 1.3 \citep[e.g.][]{1999ApJ...517..565P,2005astro.ph.10447A}. SNe~Ia
with very different stretches have been discovered locally, for
example the broad $s=1.6$ light-curves of very rare SN~2001ay-like SNe
\citep{2004howell2001ay} or the faint, fast declining SN~1991bg-like
objects \citep[e.g.][]{1992AJ....104.1543F,2004ApJ...613.1120G}.
Although SNe such as SN~1991bg have been shown to be calibrateable as
standardized candles, and have been used in determinations of the
Hubble constant \citep{2004ApJ...613.1120G}, these classes of peculiar
SNe are unlikely to be useful for cosmological studies at
high-redshift -- they may not follow the lightcurve-shape/luminosity
relationship (in the case of SN~2001ay), and may be naturally selected
against at high redshift due to their faintness (in the case of
SN~1991bg-like SNe, which are around 2 magnitudes under-luminous).  In
``distance-limited'' surveys, which likely provide a census of all
types of SNe~Ia, around 65\% of SNe~Ia appear spectroscopically
``normal'' \citep{2001ApJ...546..734L}, a number which is probably a
lower limit on the number of cosmologically useful SNe, as it excludes
SN~1991T-like events that may still be calibrateable candles.  This
fraction will only rise in high-redshift searches as the fainter
SN~1991bg-like SNe are selected against. Hence candidates with fitted
stretches outside of the range 0.7 to 1.3 are either unlikely to be
SNe~Ia, or unlikely to be cosmologically useful SNe~Ia, while those
with $0.7<s<1.3$ are likely to be useful SNe~Ia.

\begin{figure} 
\plotone{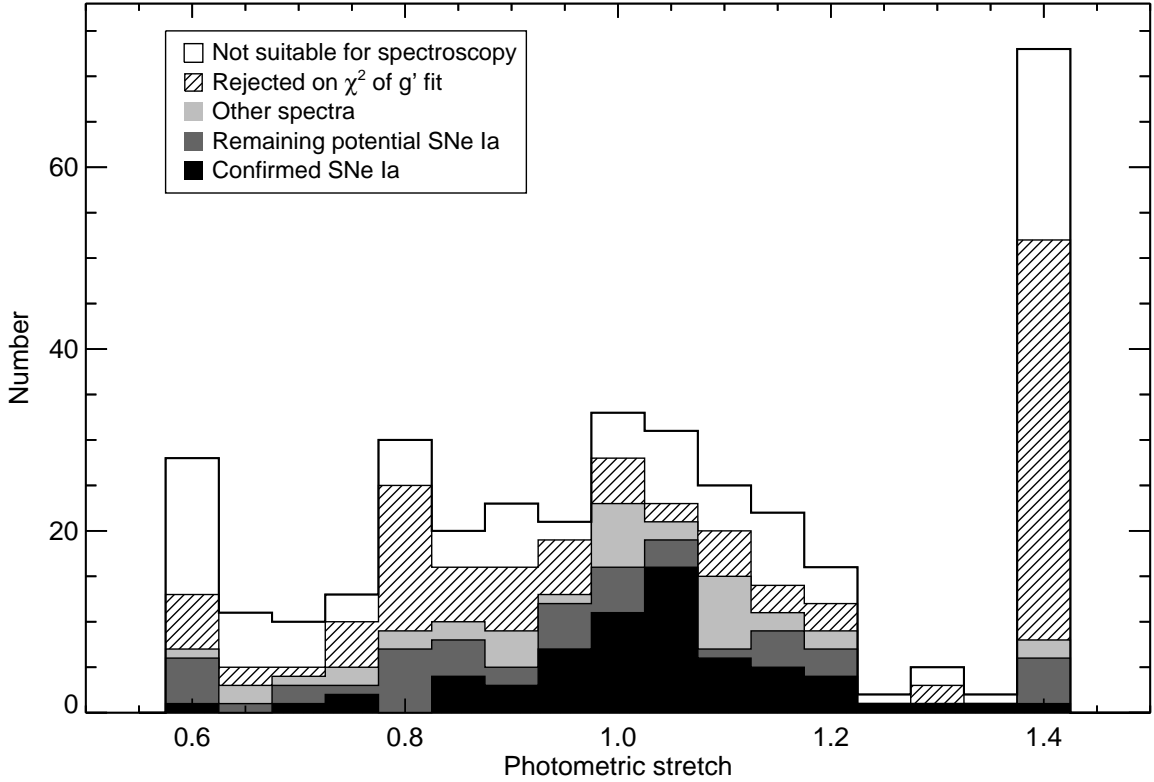}
\caption{
  The histogram of fitted stretch for all SN candidates from the SNLS
  during the time period covered by this paper.  `Confirmed SNe~Ia'
  are those SNe identified as either Ia or Ia* from their spectra,
  `Other spectra' includes any other object observed spectroscopically
  regardless of the type of object that was not identified as a SN~Ia,
  `Rejected on $\chi^2$ of $g'$ fit' includes all remaining objects
  that are rejected by the SN selection code based on the fit to the
  $g'$-band light-curve, `Not suitable for spectroscopy' includes all
  remaining objects which at maximum light are either too faint or
  have too low percentage increase for a likely successful
  spectroscopic observation, and `Remaining potential SNe' includes
  all remaining objects which are not in the above categories and
  which are \textit{potentially} SNe~Ia not observed
  spectroscopically. The photometric parameters are based on
  photometric-redshift fits to the entire candidate light-curve. The
  peaks at $s=0.6$ and $s=1.4$ reflect the range over which stretch is
  allowed to vary in the fits ($0.6<s<1.4$).
  \label{fig:stretch-hist}}
\end{figure}

\begin{figure} 
\plotone{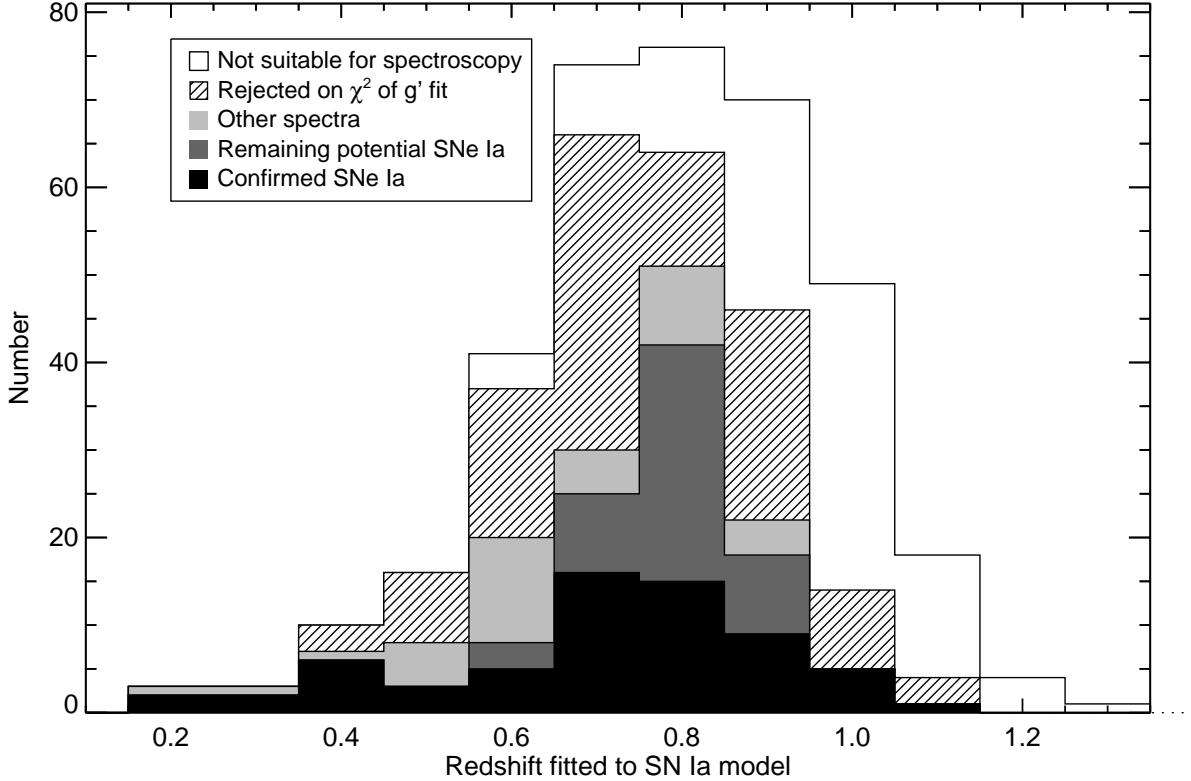}
\caption{
  The histogram of SN~Ia photometric redshifts for all candidates from
  the SNLS during the time period covered by this paper.  The
  categories of objects are the same as in
  Fig.~\ref{fig:stretch-hist}.  The photometric redshift is based on
  fits to the entire candidate light-curve under the assumption that
  the object is a SN~Ia; clearly for non-SNe~Ia the \zphot\ will be
  inaccurate. The figure shows that there are few good candidates
  lying at $\zphot<0.65$ which were rejected by the selection code in
  the real-time analysis. Note that the redshift distribution of the
  candidates `Rejected on $\chi^2$ of $g'$ fit' is not representative
  of the true redshift distribution of these events as these
  candidates are unlikely to be SNe~Ia.
  \label{fig:redshift-hist}}
\end{figure}

Figure~\ref{fig:stretch-hist} shows the distribution of fitted stretch
for all candidates (even if the spectroscopic redshift is known,
redshift is not fixed in these fits). As expected, the stretches for
the spectroscopically confirmed SNe~Ia lie in the expected range 0.7
to 1.3. Furthermore, of the candidates not selected by the code for
spectroscopic follow-up, few appear to be good candidates for normal
SNe~Ia once objects for which a spectroscopic observation would be too
challenging have been removed. Objects which were not possible to
observe spectroscopically are defined as those for which the peak $i'$
magnitude is $>24.25$, or for which the peak $i'$ percentage increase
over the host galaxy is $<30$\%. (SNe that do not meet these cuts are
rarely positively identified -- see Howell et al. 2005)
Fig.~\ref{fig:redshift-hist} shows the same objects broken-down into
the same categories, but the histogram is of the fitted redshift
rather than the stretch. As would be expected, objects defined as too
difficult to attempt spectroscopically are only found at $z>0.65$,
where the contrast over the host galaxy is typically smaller as the
hosts have a smaller apparent size on the sky - the angular separation
between SN and host is smaller, and the SN light becomes more mixed
with the galaxy light.  At $z<0.65$, of all the total SN candidates
that were consistent with a SN~Ia, only $\sim5$\% were not followed
spectroscopically where an observation would have been possible.

\section{Summary}
\label{sec:summary}

In this paper, we have presented a selection technique for
high-redshift supernovae searches that can be used to identify SNe~Ia
after only 2-3 epochs of multi-band photometry. Using both simulated
SN data and the SNLS real-time data as a test-case, we have shown that
the technique is able to discriminate between SNe~Ia and core collapse
SNe (SNe~II, SNe~Ib/c) based on the quality of the fit in the $g'$
filter. For SNe~Ia, the technique accurately predicts the redshift,
phase and light-curve parameterization of these events to a precision
of $\left|\deltaz\right|/(1+\zspec)<0.09$ in redshift, and $\pm$2-3
rest-frame days in phase, and there is no apparent bias on
cosmological parameters derived using SNe~Ia selected in using this
method.

The technique is now routinely used within the SNLS to help select
priority candidates for spectroscopic follow-up and confirmation as
SN~Ia for use as standard candles in the derivation of cosmological
parameters \citep{2005astro.ph.10447A}; the improvement in
spectroscopic success the method brings is discussed elsewhere
\citep{2005astro.ph..9195H}.  These techniques will be used in other
current and future planned surveys, which are likely to be
``spectroscopically starved'', with many more candidates than can be
followed up.

The obvious extension to this method will be when ``final'' data
(versus the real-time data used here) becomes available for SNLS. With
many of the real-time uncertainties removed, and a high quality fringe
subtraction capable of removing systematics, this SN selection
technique will provide a logical method for determining the types and
redshifts for the many hundreds of SNe that it is not possible to
observe spectroscopically, particularly when combined with photometric
redshifts for the host galaxies that will be routinely available for
the CFHT-LS Deep Fields. Such a combination of data would allow a
calculation of the rates of all SN types out to $z=1$.

\acknowledgments The SNLS collaboration gratefully acknowledges the
assistance of Pierre Martin and the CFHT Queued Service Observations
team.  Jean-Charles Cuillandre and Kanoa Withington were also
indispensable in making possible real-time data reduction at CFHT.
Based on observations obtained with MegaPrime/MegaCam, a joint project
of CFHT and CEA/DAPNIA, at the Canada-France-Hawaii Telescope which is
operated by the National Research Council (NRC) of Canada, the
Institut National des Science de l'Univers of the Centre National de
la Recherche Scientifique (CNRS) of France, and the University of
Hawaii. This work is based in part on data products produced at the
Canadian Astronomy Data Centre as part of the Canada-France-Hawaii
Telescope Legacy Survey, a collaborative project of NRC and CNRS.
Canadian collaboration members acknowledge support from NSERC and
CIAR; French collaboration members from CNRS/IN2P3, CNRS/INSU and CEA.
Based on observations (Program-IDs: GN-2004B-Q-16, GS-2004B-Q-31,
GN-2004A-Q-19, GS-2004A-Q-11, GN-2003B-Q-9, and GS-2003B-Q-8) obtained
at the Gemini Observatory, which is operated by the Association of
Universities for Research in Astronomy, Inc., under a cooperative
agreement with the NSF on behalf of the Gemini partnership: the
National Science Foundation (United States), the Particle Physics and
Astronomy Research Council (United Kingdom), the National Research
Council (Canada), CONICYT (Chile), the Australian Research Council
(Australia), CNPq (Brazil) and CONICET (Argentina).  Based on
observations made with ESO Telescopes at the Paranal Observatories
under programme ID $<$171.A-0486$>$. Some of the data presented herein
were obtained at the W.M. Keck Observatory, which is operated as a
scientific partnership among the California Institute of Technology,
the University of California and the National Aeronautics and Space
Administration. The Observatory was made possible by the generous
financial support of the W.M. Keck Foundation. The authors wish to
recognize and acknowledge the very significant cultural role and
reverence that the summit of Mauna Kea has always had within the
indigenous Hawaiian community.  We are most fortunate to have the
opportunity to conduct observations from this mountain. This research
used resources of the National Energy Research Scientific Computing
Center, which is supported by the Office of Science of the U.S.
Department of Energy under Contract No. DE-AC03-76SF00098. We thank
them for a generous allocation of computing time.

\clearpage

\begin{deluxetable}{cllc}
\tablecaption{CFHT-LS Deep field locations}
\tablehead{\colhead{Field} & \colhead{RA~(J2000)} & \colhead{DEC~(J2000)} & \colhead{Other data}}
\startdata
D1 & 02:26:00.00 & $-$04:30:00.0 & XMM-Deep, VIMOS, SWIRE, GALEX, VLA\\
D2 & 10:00:28.60 & +02:12:21.0 & COSMOS/ACS, VIMOS, SIRTF, GALEX, VLA\\
D3 & 14:19:28.01 & +52:40:41.0 & (Groth strip); DEEP-2, SIRTF, GALEX\\
D4 & 22:15:31.67 & $-$17:44:05.7 & XMM-Deep, GALEX\\
\enddata
\label{tab:fieldpositions}
\end{deluxetable}

\begin{deluxetable}{cccccc}
\tablecaption{Exposure times and epochs for a typical month in SNLS}
\tablecolumns{6}
\tablehead{\colhead{} & \colhead{Epoch 1} & \colhead{Epoch 2} & \colhead{Epoch 3} & \colhead{Epoch 4} & \colhead{Epoch 5} \\
\colhead{$\sim$Days w.r.t. new moon:} & \colhead{-8} & \colhead{-4} & \colhead{0} & \colhead{+4} & \colhead{+8}}
\startdata
$g'$\tablenotemark{a} & 5$\times$225s & \nodata & 5$\times$225s & 5$\times$225s & \nodata \\
$r'$ & 5$\times$300s & 5$\times$300s & 5$\times$300s & 5$\times$300s & 5$\times$300s \\
$i'$ & 7$\times$520s & 5$\times$360s & 7$\times$520s & 5$\times$360s & 7$\times$520s \\
$z'$ & 10$\times$360s & \nodata & 10$\times$360s & \nodata & 10$\times$360s \\
\enddata
\tablenotetext{a}{For the $g'$ filter, epochs 1/2 and epochs 4/5 can be exchanged depending on moon position. From August 2005, 4 $g'$ epochs have been obtained.}
\label{tab:exposure_times}
\end{deluxetable}

\clearpage 

\begin{deluxetable}{ccccc}
\tablecaption{``actual minus recovered'' SN parameters for the 2000 synthetic SNe~Ia}
\tablecolumns{5}
\tablehead{\colhead{Parameter} & \multicolumn{2}{c}{Simple} & \multicolumn{2}{c}{Outlier resistant}\\ \colhead{} & \colhead{Mean} & \colhead{$\sigma$} & \colhead{Mean} & \colhead{$\sigma$} }
\startdata
\sidehead{Pre-maximum light-curve epochs only:}
$z-z_{\mathrm{recovered}}$&0.098&0.240&0.007&0.025\\
$s-s_{\mathrm{recovered}}$&0.204&0.423&0.015&0.131\\
$\mathrm{d}m-\mathrm{d}m_{\mathrm{recovered}}$&-0.004&0.144&-0.004&0.140\\
\sidehead{Entire light-curve:}
$z-z_{\mathrm{recovered}}$&0.006&0.041&0.007&0.022\\
$s-s_{\mathrm{recovered}}$&0.041&0.099&0.021&0.044\\
$\mathrm{d}m-\mathrm{d}m_{\mathrm{recovered}}$&0.010&0.142&0.012&0.132\\
\enddata
\label{tab:fakesne_dist}
\end{deluxetable}

\clearpage 

\clearpage

\clearpage

\clearpage

\clearpage

\clearpage

\clearpage

\clearpage

\clearpage

\clearpage

\clearpage

\clearpage

\clearpage

\clearpage

\clearpage

\end{document}